\documentclass[12pt]{article}
\usepackage{newtxtext,newtxmath}
\usepackage{graphicx}
\graphicspath{{Figures/}}

\usepackage[letterpaper,margin=1in]{geometry}

\renewenvironment{abstract}
	{\quotation}
	{\endquotation}

\date{}

\makeatletter
\renewcommand{\fnum@figure}{\textbf{Figure \thefigure}}
\renewcommand{\fnum@table}{\textbf{Table \thetable}}
\makeatother

\usepackage{cuted}

\usepackage{capt-of}

\usepackage{url}

\newcommand{\kbt}{\text{k}_\text{B}\text{T}}

\usepackage{cuted} 

\usepackage{capt-of} 

\makeatletter
\g@addto@macro{\UrlBreaks}{%
  \do\/\do-\do\_\do\.\do\?\do\&\do\=\do\#\do\%\do\+%
}
\makeatother

\usepackage[colorlinks=true, allcolors=blue, breaklinks=true]{hyperref}
    
\AtBeginDocument{
  \small            
  \twocolumn        
}

\def\scititle{Pathway variability, coat stiffening and mechanical adaptation
during clathrin-mediated endocytosis}

\title{\bfseries \boldmath \scititle}

\author{
	J. H. H. Dreckhoff,
    U. S. Schwarz$^{\ast}$,
	L. Lettermann$^{\ast}$\and
	\small BioQuant, Heidelberg University, 69120 Heidelberg, Germany.\and
    \small Institute for Theoretical Physics, Heidelberg University, 69120 Heidelberg, Germany.\and
	\small$^\ast$Corresponding authors. Emails: schwarz@thphys.uni-heidelberg.de and lettermann@uni-heidelberg.de\and
}

\begin{document} 

\begin{strip}
    
\maketitle

\begin{abstract} \bfseries \boldmath
Clathrin assemblies in cells can persist as flat plaques, abort after partial invagination, 
or close into clathrin-coated vesicles, but the determinants of these different fates remain unresolved. 
To investigate the stochastic and complex dynamics of clathrin
assemblies, we have developed a kinetic Monte Carlo simulation framework that
couples individual clathrin agents to an adaptive continuum membrane. 
In this hybrid discrete-continuum description, the effective coat bending rigidity and the 
preferred coat curvature emerge during growth, rather than being prescribed as material parameters. 
Once connected, curved lattices stiffen from molecular bending modes to coat-level rigidities, 
because curvature changes require increased stretching or compression, 
while newly incorporated triskelia hardcode a history-dependent preferred curvature. 
An analytical theory for non-Euclidean elasticity identifies the relevant internal variables 
and predicts growth laws that are validated by the simulations.
The same microscopic assembly rules yield flat, stalled, 
and closed coats through two sequential gates in the effective membrane–coat energy landscape. 
Comparisons with experimentally observed coat geometries and nanodissection-induced curvature 
changes agree with our theoretical predictions without any fitting parameters. 
The clathrin coat thus emerges as an adaptive assembly with prestress and memory, 
whose fate and material parameters reflect the environment in which it has been growing.
\end{abstract}

\end{strip}

\noindent Clathrin-mediated endocytosis (CME) is the major route by which eukaryotic
cells internalize membrane receptors, nutrients, and signaling molecules.~
\cite{andersonRoleCoatedEndocytic1977,andersonMutationThatImpairs1977,
goldsteinCoatedPitsCoated1979,fotinMolecularModelComplete2004,
hauckeMembraneRemodelingClathrinmediated2018,
kaksonenMechanismsClathrinmediatedEndocytosis2018} During CME, clathrin
triskelia assemble into a polygonal lattice coat that remodels the membrane
first into a clathrin-coated pit (CCP) and then into a clathrin-coated
vesicle (CCV).~\cite{pearseCoatedVesiclesPig1975, pearseClathrinUniqueProtein1976,
crowtherAssemblyPackingClathrin1981,ungewickellAssemblyUnitsCIathrin1981,
sochackiStructureSpontaneousCurvature2021,morrisCryoEMMultipleCage2019,
mcmahonMolecularMechanismPhysiological2011,
brodskyDiversityClathrinFunction2012,
kirchhausenMolecularStructureFunction2014} Mature coats are far stiffer than
individual triskelion legs: measured effective coat rigidities are of order
$10^2$--$10^3\,\kbt$, whereas molecular leg bending costs are of order
$20\,\kbt$.~\cite{jinRigidityTriskelionArms2000,
nossalEnergeticsClathrinBasket2001,jinMeasuringElasticityClathrinCoated2006,
tagiltsevNanodissectedElasticallyLoaded2021} 
How this mechanical stiffening emerges from the growing lattice and what ultimately drives curvature generation remain unclear,
including the question whether pentagonal defects are a cause or a consequence of membrane bending.

\begin{figure*}[!htbp]
	\centering
	\includegraphics[width=1.0\textwidth, angle=0]{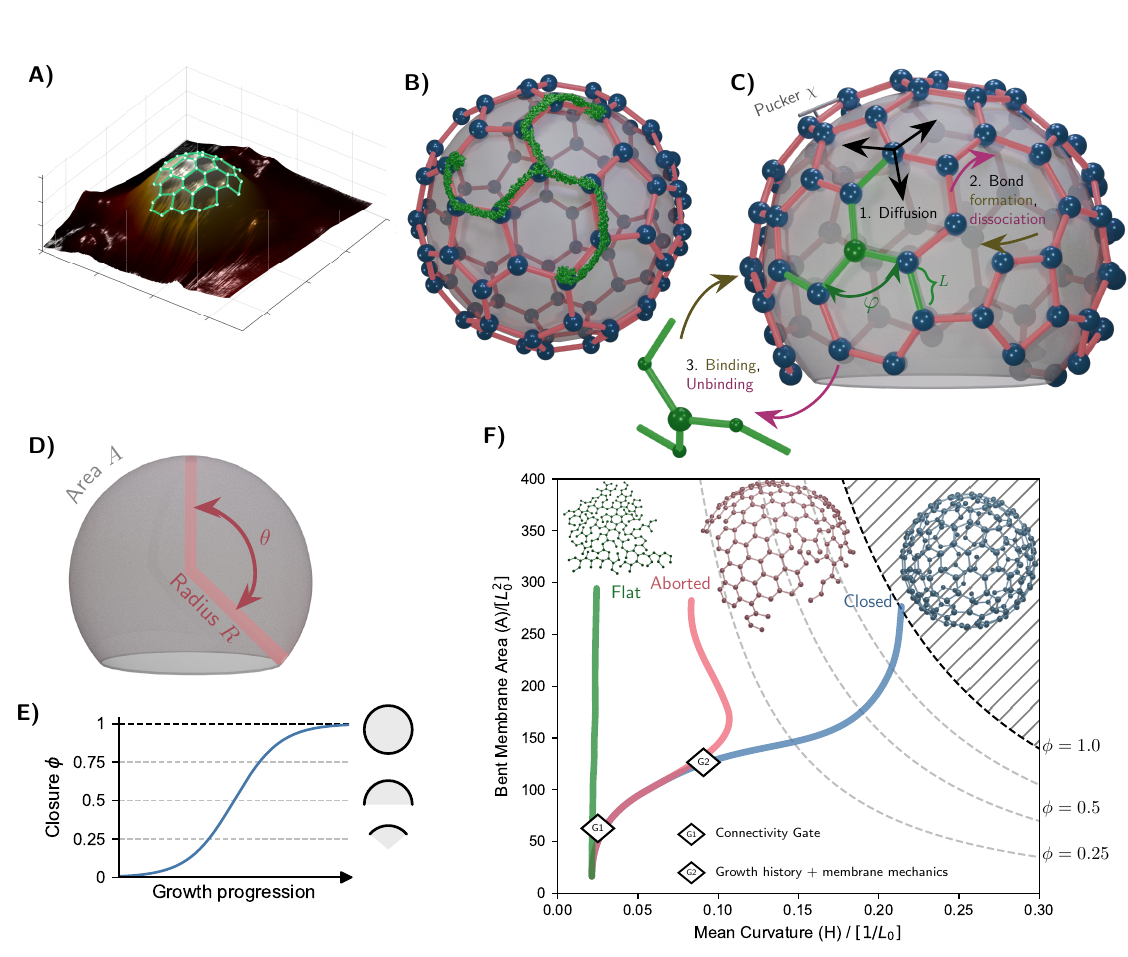}

	\caption{
    \textbf{Kinetic Monte Carlo computer simulations 
    explain complex dynamics of clathrin assemblies.}
		(\textbf{A}) Three-dimensional rendering of a membrane-associated
        clathrin coat reconstructed from high-speed AFM data, with individual
        clathrin--clathrin connections highlighted in green. Image reproduced from~\cite{tagiltsevNanodissectedElasticallyLoaded2021} with permission. (\textbf{B})
        Representative simulated coat with a molecular clathrin structure from
        the Protein Data Bank overlaid onto the lattice. In the
        simulation, triskelia are represented as discrete mobile nodes
        connected by deformable legs. (\textbf{C}) Microscopic
        degrees of freedom and stochastic moves: bound nodes diffuse on the
        membrane, clathrin--clathrin bonds form and dissociate, and triskelia
        bind to or unbind from the membrane. The single clathrin state is
        parameterized by leg length $L$, in-plane angle $\varphi$, and pucker
        angle $\chi$; binding, unbinding, and bond remodeling are treated as
        kinetic Monte Carlo events while membrane curvature adapts to the
        assembly state. (\textbf{D}) Spherical-cap
        geometry used to couple the two-dimensional coat to membrane
        curvature, described by area $A$, radius (inverse curvature) $R=1/H$, opening angle
        $\theta$, and closure $\phi=AH^2/4\pi$. (\textbf{E}) Closure $\phi$ measures the occupied membrane area compared to the total area of a sphere of equal curvature, and will increase for a generic growth process. (\textbf{F}) Area--curvature representation of the
        three simulated assembly fates: flat plaques (green), aborted pits
        (pink), and closed coats (blue). Dashed contours indicate lines of constant
        closure, the shaded region is inaccessible as it lies beyond complete spherical closure
        ($\phi>1$), and example simulated coats illustrate the final
        morphologies. The two fate gates, deciding over the evolution branch, are shown as diamonds.}
	\label{fig:ModelExplanation} 
\end{figure*}

Because of their small size, clathrin structures have long been imaged only by electron microscopy. Coated invaginations first appeared as ``bristle-coated'' pits in thin-section micrographs~\cite{rothYOLKPROTEINUPTAKE1964}, followed by the characteristic polygonal basket of pentagons and hexagons~\cite{kanasekiVESICLEBASKET1969}; the coat protein was then isolated and named clathrin~\cite{pearseClathrinUniqueProtein1976}. Deep-etch electron microscopy gave the first three-dimensional views of coated-pit formation~\cite{heuserThreedimensionalVisualizationCoated1980}, cryo-electron microscopy a pseudo-atomic model of the complete lattice~\cite{fotinMolecularModelComplete2004}, and platinum-replica EM and cryo-electron tomography mapped flat and curved lattices directly at the plasma membrane~\cite{sochackiStructureSpontaneousCurvature2021,morrisCryoEMMultipleCage2019}. More recently, correlated light and electron microscopy~\cite{bucherClathrinadaptorRatioMembrane2018}, super-resolution microscopy~\cite{mundClathrinCoatsPartially2023}, and atomic force microscopy~\cite{tagiltsevNanodissectedElasticallyLoaded2021} have linked lattice state to assembly progress at increasing resolution; Fig.~\ref{fig:ModelExplanation}A shows an AFM-reconstructed coat from which the lattice geometry can be read off. Single triskelia and the resulting lattice geometry are thus well characterized (Fig.~\ref{fig:ModelExplanation}B), yet following one coat as it grows in molecular detail remains out of reach. Structural methods require fixed samples and cannot track a single coat over time~\cite{heuserThreedimensionalVisualizationCoated1980,sochackiStructureSpontaneousCurvature2021,mundClathrinCoatsPartially2023}; pseudotime ordering of many fixed snapshots approximates an average trajectory~\cite{mundClathrinCoatsPartially2023}, but mixes different pathways to final assembly and blurs their differences. Live-cell fluorescence, conversely, follows single sites across their lifetime but resolves neither the molecular lattice nor local curvature or individual binding events~\cite{kirchhausenImagingEndocyticClathrin2009}.

In cells, not every assembly reaches closure: coats are found as persistent flat plaques, aborted pits, or closed vesicles, and what sets these fates remains unclear~\cite{sochackiStructureSpontaneousCurvature2021,mundClathrinCoatsPartially2023,bucherClathrinadaptorRatioMembrane2018,willyNovoEndocyticClathrin2021}. Existing models have treated limiting scenarios such as constant-curvature~\cite{kirchhausenCoatedPitsCoated1993} or constant-area~\cite{heuserThreedimensionalVisualizationCoated1980} growth, and energy-landscape descriptions show how membrane tension and polymerization bias invagination pathways~\cite{avinoamEndocyticSitesMature2015,mundClathrinCoatsPartially2023,saleemBalanceMembraneElasticity2015,freyCompetingPathwaysInvagination2020}. These frameworks capture much of CME phenomenology, but typically treat coat stiffness and preferred curvature as fixed material parameters rather than as quantities generated by assembly itself. Coat rigidity has recently been proposed to increase dynamically during growth~\cite{freyCoatStiffeningCan2024a}, yet a mechanistic framework that bridges microscopic lattice assembly and continuum membrane deformation 
with dynamical parameters is still missing.

This missing bridge reflects the mixed nature of the problem: membrane bending is a continuum elastic process~\cite{helfrichElasticPropertiesLipid1973}, whereas clathrin assembly is discrete, with individual triskelia binding, diffusing, forming local contacts, leaving vacancies and defects, and ultimately satisfying the topology of a curved trivalent lattice (Fig.~\ref{fig:ModelExplanation}C)~\cite{jinTopologicalMechanismsInvolved1993,morrisCryoEMMultipleCage2019,denotterGenerationCurvedClathrin2011, guo2022large}. Since this regime is inaccessible to live-cell imaging, it is a natural target for computer simulation~\cite{denotterGenerationCurvedClathrin2011}. Earlier simulations, however, imposed coat curvature through prescribed conformational switching of individual triskelia rather than letting it emerge from the assembly energetics, and did not address dynamic coat stiffening, history-dependent curvature memory, or the mechanical selection between flat, aborted, and closed fates.

Here we introduce a hybrid kinetic Monte Carlo model that couples an
agent-based clathrin coat to a continuum membrane geometry. Each
triskelion is represented as an individual mobile agent on a space-continuous membrane, inspired by the molecular structure (Fig.~\ref{fig:ModelExplanation}B). Its
three legs are coarse-grained into the hub-to-hub-bearing segments that form
the lattice. Each agent carries a
local elastic Hamiltonian with costs for deviations from the preferred leg
length $L_0$, in-plane angle $\varphi_0$ and pucker angle $\chi_0$, encoding
lateral lattice organization and microscopic curvature preference
(Fig.~\ref{fig:ModelExplanation}C). Nodes diffuse on the membrane, bind and
unbind, and form or dissolve clathrin--clathrin contacts with
energy-dependent stochastic rates. The hub-to-hub length
$L_0=18.4\,\mathrm{nm}$ sets the length unit throughout.

Membrane shape enters through a two-dimensional spherical-cap geometry, described by coat area $A$ and mean curvature $H$, or, equivalently, by the invagination angle $\cos\theta=1-AH^2/2\pi\in[1,-1]$ (Fig.~\ref{fig:ModelExplanation}D). In the following, we use the more convenient closure variable $\phi=AH^2/4\pi\in[0,1]$, which is the ratio of occupied membrane area to the area of a sphere of equal curvature. It starts at $\phi=0$, rises sharply during growth, and plateaus at $\phi=1$ (Fig.~\ref{fig:ModelExplanation}E). The molecular degrees of freedom remain surface coordinates, but the surface on which they move can bend, and between stochastic assembly events the membrane curvature adapts adiabatically to the current mechanical state of the coat. The model thereby gives access to the time-resolved microstructural dynamics of coat assembly in a single dynamical framework.

As we will show in the following, from this single model three assembly fates emerge: flat plaques, aborted pits, and closed CCVs, matching the main morphological classes seen experimentally (Fig.~\ref{fig:ModelExplanation}F), where fate selection proceeds through two sequential gates set by the interplay of discrete and continuum effects. The first gate (G1) is lattice connectivity: coats too poorly connected to transmit coherent bending stresses remain flat, consistent with loose or vacancy-rich flat lattices observed or inferred in cells~\cite{sochackiStructureSpontaneousCurvature2021,freyEdenGrowthModels2020}, whereas sufficiently connected coats generate curvature and stiffen through curved spherical assembly. The second gate (G2) is rooted in the continuum energy landscape: for curvature-generating coats the average closure is set by mechanical parameters such as membrane tension, polymerization energy, and membrane bending rigidity, with the surface-tension dependence the clearest signature in the simulations. 
We find that topological defects such as pentagons and heptagons are not the primary trigger of curvature; rather they follow and stabilize it, reconciling the mechanical pathway with the topological requirement for closed clathrin cages~\cite{jinTopologicalMechanismsInvolved1993,sochackiStructureSpontaneousCurvature2021,morrisCryoEMMultipleCage2019}. Combining the simulations with a theory for non-Euclidean elasticity that captures memory and prestress, we identify the underlying growth laws and the precise nature of these gates, and thereby explain the observed pathway variability with one set of physical rules.

\begin{figure*}
	\centering
	\includegraphics[width=1.0\textwidth, angle=0]{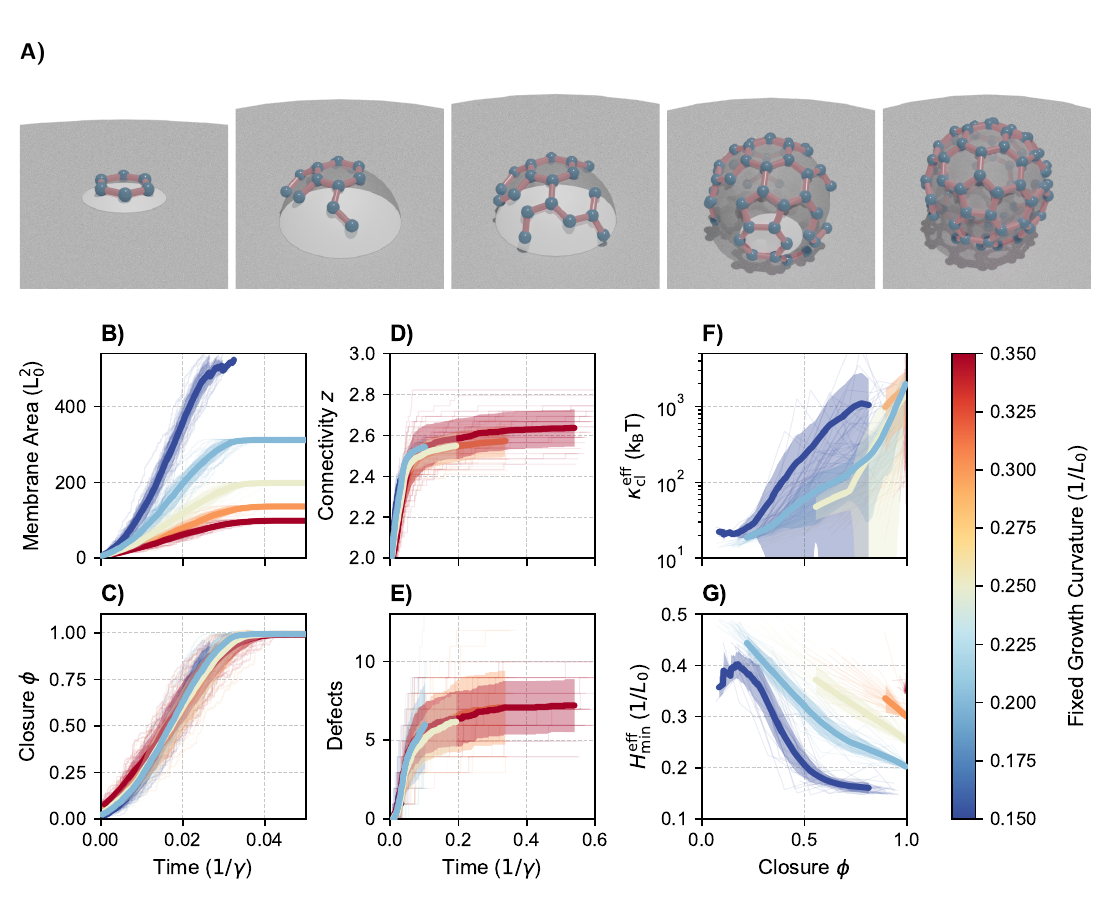} 
	\caption{\textbf{Structural assembly at fixed growth curvature reveals
    dynamical stiffening and curvature memory.}
		(\textbf{A}) Snapshots of a clathrin lattice growing on a spherical cap
        with fixed membrane curvature, from early assembly to closure. The
        membrane is shown in gray, clathrin hubs in blue, and clathrin legs in
        red. See also Supplementary Movies S1 and S2. (\textbf{B}-\textbf{G}) Growth metrics for ensembles assembled at
        imposed curvatures from $0.15\,L_0^{-1}$ to $0.35\,L_0^{-1}$, indicated
        by color. Thin lines show individual trajectories, thick lines show
        ensemble means, and shaded regions indicate one standard deviation.
        Time is given in units of the inverse assembly rate, $\gamma^{-1}$. (\textbf{B}) and (\textbf{C}) focus on the early evolution.
        (\textbf{B}) Clathrin-covered membrane area. Lower-curvature coats
        occupy larger final areas because the corresponding target sphere is
        larger. (\textbf{C}) Geometric closure
        $\phi=AH^2/4\pi$, with $\phi=1$ marking a closed spherical coat.
        (\textbf{D}) Mean connectivity $z$, defined as the
        average number of clathrin--clathrin bonds per hub. (\textbf{E})
        Total number of pentagons plus heptagons, $P+H$, used as a
        defect-load measure. (\textbf{F})
        Effective coat bending rigidity $\kappa_{\rm cl}^{\rm eff}$ plotted
        against closure on a logarithmic scale. (\textbf{G}) Effective
        energy-minimizing curvature $H_{\rm min}^{\rm eff}$ plotted against closure,
        showing the shift of the preferred coat curvature during growth, with all trajectories ending close to their imposed growth curvature.}
	\label{fig:FixedCurvature} 
\end{figure*}

\subsection*{Clathrin coats stiffen and develop curvature memory during assembly}

We first ask how the mechanical state of a coat changes while the lattice
assembles. To isolate lattice mechanics from membrane-shape dynamics, we first
grow clathrin at prescribed membrane curvature. This fixed-curvature simulation is
experimentally motivated by clathrin assembly on curved membrane templates and
by particle-associated uptake geometries, where clathrin and adaptor assemblies
respond strongly to local membrane curvature~\cite{
zenoClathrinSensesMembrane2021,voigtSubstrateStiffnessParticle2024}. In the
simulation, individual triskelia bind, unbind, diffuse, and remodel bonds on a
spherical cap whose curvature is held fixed. The coat therefore assembles on a
curved surface, but it does not yet have to bend the membrane itself.
Figure~\ref{fig:FixedCurvature}A shows that compact, ordered trivalent coats
emerge from local triskelion rules without imposing a grid, polygon template,
or cage architecture.

Figures~\ref{fig:FixedCurvature}B--G quantify ensembles grown across imposed
curvatures from $0.15$ to $0.35\,L_0^{-1}$. We track growth both by time, in
units of the inverse microscopic assembly rate $\gamma^{-1}$, and by the
geometric closure variable $\phi=AH^2/4\pi$, where $A$ is the clathrin-covered
area and $H$ is the membrane curvature. Lower-curvature coats require more
absolute area to close because their target sphere is larger
(Fig.~\ref{fig:FixedCurvature}B). These coats also grow faster, because
they have a larger rim at which new triskelia can dock, in agreement
with reaction-limited Eden-type growth models~\cite{freyEdenGrowthModels2020}.
Rescaled to closure instead of absolute area, trajectories
have the same sigmoidal shape for all curvatures
(Fig.~\ref{fig:FixedCurvature}C).

The fixed-curvature simulations also reveal how lattice connectivity evolves when
the coat does not have to bend the membrane itself. The mean connectivity $z$, or coordination number,
is the average number of clathrin--clathrin bonds per hub
(Fig.~\ref{fig:FixedCurvature}D).
Across imposed curvatures, connectivity follows similar
trajectories, although the largest, low-curvature coats remain slightly less
connected on average. A natural explanation is geometric crowding: on small,
high-curvature caps, triskelia have less lateral space and more readily form
nearby contacts, whereas on larger, low-curvature caps, they can assemble more
sparsely while still completing the externally prescribed spherical geometry.

Defects provide the corresponding topological readout
(Fig.~\ref{fig:FixedCurvature}E). For an ideal closed trivalent cage on a
sphere, Euler topology requires twelve more pentagons than heptagons,
$P-H=12$~\cite{jinTopologicalMechanismsInvolved1993,
bowickTwoDimensionalMatterOrder2009}. During assembly, the coat is open, has a
boundary, and may contain missing bonds, so the closed-cage constraint need not
hold at intermediate stages. We therefore plot the total defect load $P+H$,
which counts pentagons and heptagons without identifying them with the
topological charge. In the fixed-curvature simulation, this defect load rises with
time and then plateaus with only weak dependence on imposed curvature, 
showing that the structural maturation of the lattice is broadly similar
across the curvature range.

The mechanical response changes far more strongly. For each assembled coat, we
compute an effective coat bending rigidity $\kappa_{\rm cl}^{\rm eff}$ by copying
the configuration, perturbing its curvature, measuring the energy response,
and fitting that response to a Helfrich-type curvature energy.~\cite{helfrichElasticPropertiesLipid1973} Because the microscopic parameters are fixed throughout
the simulations, changes in $\kappa_{\rm cl}^{\rm eff}$ report changes generated by
assembly itself. As an overall trend, we observe that at small closure, the 
coat is compliant, with
$\kappa_{\rm cl}^{\rm eff}$ of order $20\,\kbt$, comparable to the microscopic
leg-bending scale $k_{\rm bend}\sim 20\,\kbt$ listed in Table~\ref{tab:key parameters} (detailed derivation of microscopic constants can be found in Supplementary Note 1.1). During growth, the
effective coat bending rigidity increases by almost two orders of magnitude and reaches
values of order $10^3\,\kbt$ near closure
(Fig.~\ref{fig:FixedCurvature}F), placing mature simulated coats in the same
broad mechanical regime as experimentally inferred clathrin-coated
structures~\cite{jinRigidityTriskelionArms2000,
nossalEnergeticsClathrinBasket2001,jinMeasuringElasticityClathrinCoated2006,
tagiltsevNanodissectedElasticallyLoaded2021}. The final values approach the
microscopic stretching scale $k_{\rm stretch}\sim800\,\kbt$, indicating that
the connected coat progressively converts cheap angular rearrangements into
costly stretching-dominated resistance. The similar stiffening trajectories
across imposed curvatures suggest a common geometric growth law, which we will
identify later. Ensemble averages for fitted mechanical quantities are shown only over
closure intervals with enough configurations for reliable curvature-response
fits; high-curvature coats reach large closure rapidly and therefore
contribute shorter reliable fit ranges.

Assembly also changes the curvature that minimizes the overall coat energy.
From the same curvature-perturbation fits, we obtain the effective
energy-minimizing curvature $H_{\rm min}^{\rm eff}$
(Fig.~\ref{fig:FixedCurvature}G). If this preference were set only by
individual triskelion geometry, it would remain close to the microscopic
energy-minimizing curvature $H_{\rm min}^{\rm micro}\simeq0.5\,L_0^{-1}$,
calibrated from clathrin cages assembled in vitro without membrane present~\cite{
crowtherAssemblyPackingClathrin1981,vigersThreedimensionalStructureClathrin1986,
nossalEnergeticsClathrinBasket2001,morrisCryoEMMultipleCage2019}. Instead,
$H_{\rm min}^{\rm eff}$ shifts during growth toward the imposed growth
curvature $H_{\rm growth}$, and the averaged trajectories end close to their
respective imposed curvatures. A closed coat grown at fixed curvature has
therefore adopted that curvature as its preferred mechanical state.

We interpret this shift as curvature memory. Each newly incorporated
triskelion is added in a relaxed configuration at the current membrane
curvature and thereby writes a local reference geometry into the growing coat.
This is the clathrin analogue of elastic memory in non-Euclidean sheets, where
growth or assembly defines local rest distances that need not match a later
shape~\cite{efratiElasticTheoryUnconstrained2009,
sharonMechanicsNonEuclideanPlates2010,meiriCumulativeGeometricFrustration2021,
kleinShapingElasticSheets2007}. 
Fixed-curvature growth thus establishes two
assembly-generated state variables, the effective coat bending rigidity
$\kappa_{\rm cl}^{\rm eff}$ and the effective energy-minimizing curvature
$H_{\rm min}^{\rm eff}$, which we next allow to feed back onto membrane shape, releasing the previous fixed curvature constraint.

\begin{strip}
\begin{center}
    \centering
	\includegraphics[width=1.0\textwidth, angle=0]{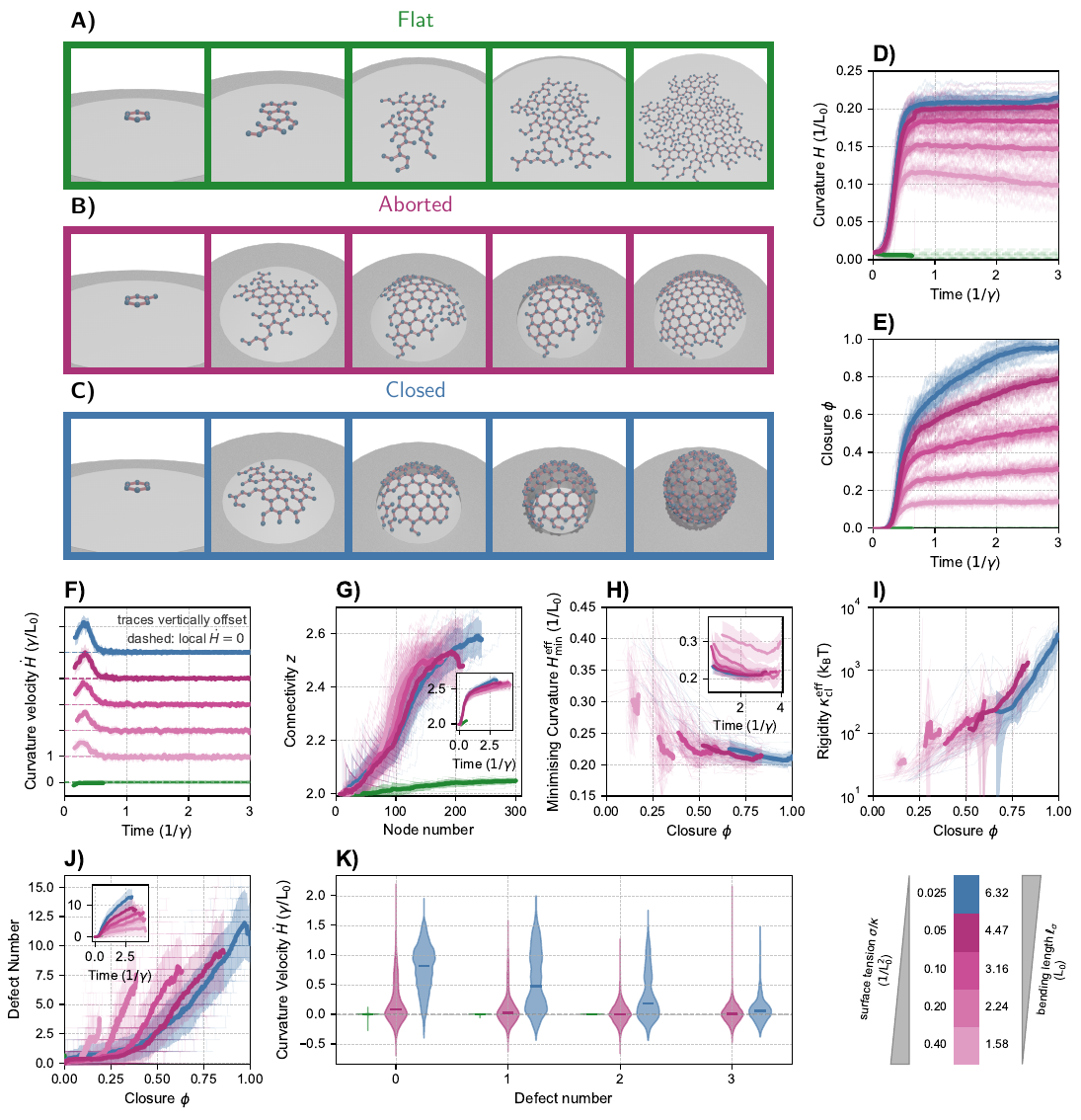} 
\end{center}
\captionof{figure}{\textbf{Growth on a deformable membrane produces three assembly fates.}
        (\textbf{A}-\textbf{C}) Representative trajectories of flat, aborted,
        and closed coats generated by the same microscopic clathrin model. See also Supplementary Movies S3 -- S5.
        (\textbf{D}-\textbf{K}) Growth metrics for assembled coat ensembles with variable
        curvature during growth. Color indicates final fate, and the pink and blue color
        scale indicates the membrane bending length
        $\ell_\sigma=\sqrt{\kappa_{\rm M}/\sigma}$ or surface tension ratio $\sigma/\kappa_{\rm M}$, with
        membrane bending rigidity $\kappa_{\rm M}$ and membrane tension $\sigma$.
        (\textbf{D}, \textbf{E}) Membrane curvature and closure
        during growth show that curvature-generating coats undergo a pronounced
        bending phase followed by slower evolution; only closed trajectories
        reach $\phi=1$. (\textbf{F}) Curvature generation as change in curvature per
        unit time has a pronounced peak during early development. Different batches shifted in y-direction. (\textbf{G}) Mean connectivity 
        $z$ plotted against the number of
        incorporated clathrin hubs separates flat plaques from
        curvature-generating coats, indicating the first fate gate. The inset shows the corresponding time
        traces. (\textbf{H}) Effective energy-minimizing curvature
        $H_{\rm min}^{\rm eff}$ shows pronounced decrease with closure. The inset shows
        the corresponding time evolution. (\textbf{I}) Effective bending rigidity
        $\kappa_{\rm cl}^{\rm eff}$ again shows increase during growth with closure.
         (\textbf{J}) Defect number plotted against
        closure. (\textbf{K}) Change in membrane curvature per unit simulation
        time plotted against defect number, showing that curvature can be
        generated before visible topological defects are present.}
	\label{fig:VariableCurvature} 
\end{strip}

\subsection*{Three assembly fates emerge from a minimal model}

We next allow membrane shape to evolve with the assembling coat. 
The simulation now starts from a flat
membrane, so curvature is no longer prescribed, but must be produced
by the growing lattice itself. Between stochastic clathrin binding, unbinding,
diffusion, and bond changes, the spherical-cap curvature adiabatically
adapts its mechanical state. We keep the
microscopic clathrin parameters fixed and vary the membrane bending length
$\ell_\sigma=\sqrt{\kappa_{\rm M}/\sigma}$ which sets the scale at which membrane
bending and surface-tension costs balance, by varying the surface tension $\sigma$ 
and keeping the membrane bending rigidity $\kappa_{\rm M}$ fixed. 
This choice reflects the established
role of membrane mechanics in clathrin-coated pit shape and invagination
efficiency~\cite{saleemBalanceMembraneElasticity2015,
bucherClathrinadaptorRatioMembrane2018,hassingerDesignPrinciplesRobust2017}.
We note that in our framework surface tension $\sigma$ is the main
parameter representing the role of the environment on CME, but that other
environmental features might also play a role, like adhesion and growth
of actin structures.

With the same microscopic assembly rules, the model produces three outcomes:
flat plaques, aborted pits, and closed coats
(Fig.~\ref{fig:VariableCurvature}A--C). Importantly, fate is not determined
only by the external parameter set. Although the ensemble averages separate
clearly with imposed bending length $\ell_\sigma$, individual trajectories can reach
different outcomes under the same parameters, because their stochastic assembly
histories generate different connectivities, growth curvatures, and mechanical
states. In this sense, pathway variability is already present at the level of
the minimal lattice dynamics.

Flat plaques grow in area but do not generate appreciable curvature. Aborted
pits generate curvature but do not complete vesicle closure. Biologically,
such abortive outcomes can arise when closure is too slow to occur before
coat disassembly, or when the coat reaches a configuration in which further
closure is mechanically unfavorable. The finite simulated trajectory length is
therefore used only as an operational proxy for the finite cellular lifetime of
clathrin-coated pits~\cite{loerkeCargoDynaminRegulate2009,
saffarianDistinctDynamicsEndocytic2009,cocucciFirstFiveSeconds2012, guo2022large}. Closed
coats, in contrast, reach the geometric closure condition $\phi=1$. The
energy-landscape analysis below will separate slowly closing trajectories from
trajectories that stall at an internal mechanical state.

The time traces show how the fates separate dynamically
(Fig.~\ref{fig:VariableCurvature}D, E, F).
After brief flat growth, curvature-generating coats undergo
an initial bending phase, visible as a rapid increase in curvature and closure.
Together with a transient peak in the change in curvature
per unit simulation time, this represents an emergent flat-to-curved transition. 
This is followed by a slower phase in which
curvature changes little and additional closure is dominated by area growth.
The bending length modulates this progression: mechanically resistant membranes
show slower curvature generation and lower final closure.

The first structural discriminator is lattice connectivity
(Fig.~\ref{fig:VariableCurvature}G). Plotting the mean connectivity $z$ against
the number of incorporated clathrin hubs shows that coats that bend the
membrane progressively incorporate hubs with higher connectivity, whereas flat
plaques remain sparsely connected as they grow. This suggests a connectivity gate: a
coat must become sufficiently connected to transmit bending stresses to the
membrane. The result is consistent with experimental and theoretical pictures
of loose or vacancy-rich flat clathrin lattices~\cite{
sochackiStructureSpontaneousCurvature2021,freyEdenGrowthModels2020,
broederszModelingSemiflexiblePolymer2014}. Because live structural data can
show proximity of clathrin legs more directly than mechanical bond strength,
we do not equate the simulated value of $z$ with a directly measured cellular
quantity. Instead, the model result identifies a necessary mechanical
condition: a lattice with too little stress transmission remains flat. Other
cellular mechanisms may also weaken or strengthen stress transmission, but
some sufficient level of mechanical coupling is required for the
geometry-induced stiffening derived below.

Topological defects are not required to initiate curvature generation
(Fig.~\ref{fig:VariableCurvature}J, K). 
A closed trivalent cage must ultimately
satisfy the Euler constraint of twelve excess pentagons, but the simulations
show that curvature can start before visible non-hexagonal polygons are
counted. Closed trajectories already display large changes in membrane
curvature per unit simulation time at zero counted defects. Defects therefore
follow and may stabilize curvature generated by a sufficiently connected coat,
rather than acting as the primary mechanical trigger~\cite{
jinTopologicalMechanismsInvolved1993,bowickTwoDimensionalMatterOrder2009}.

The mechanical state variables identified in the fixed-curvature assay remain
dynamic on a deformable membrane. The effective energy-minimizing curvature
$H_{\rm min}^{\rm eff}$ shifts during growth from the microscopic
preferred curvature toward the curvature realized by the coat
(Fig.~\ref{fig:VariableCurvature}H), while the effective coat bending rigidity
$\kappa_{\rm cl}^{\rm eff}$ increases strongly with closure
(Fig.~\ref{fig:VariableCurvature}I). The averages are shorter than some
individual trajectories, because reliable curvature-perturbation fits require
enough data points in each fit range, but the qualitative result matches the
controlled fixed curvature assay: assembly shifts the effective energy-minimizing curvature of
the coat and progressively increases its resistance to curvature changes.

Growth on a deformable membrane therefore converts assembly-written mechanics
into fate selection. Low-connectivity coats remain flat; sufficiently connected
coats generate curvature; and among these bending trajectories, membrane
mechanical load and stochastic growth history determine
whether the trajectory stalls before closure or reaches a closed vesicle. We
next derive the geometric origin of the stiffening and curvature-memory
variables that underlie this branching.

\begin{figure*}
	\centering
	\includegraphics[width=1.0\textwidth, angle=0]{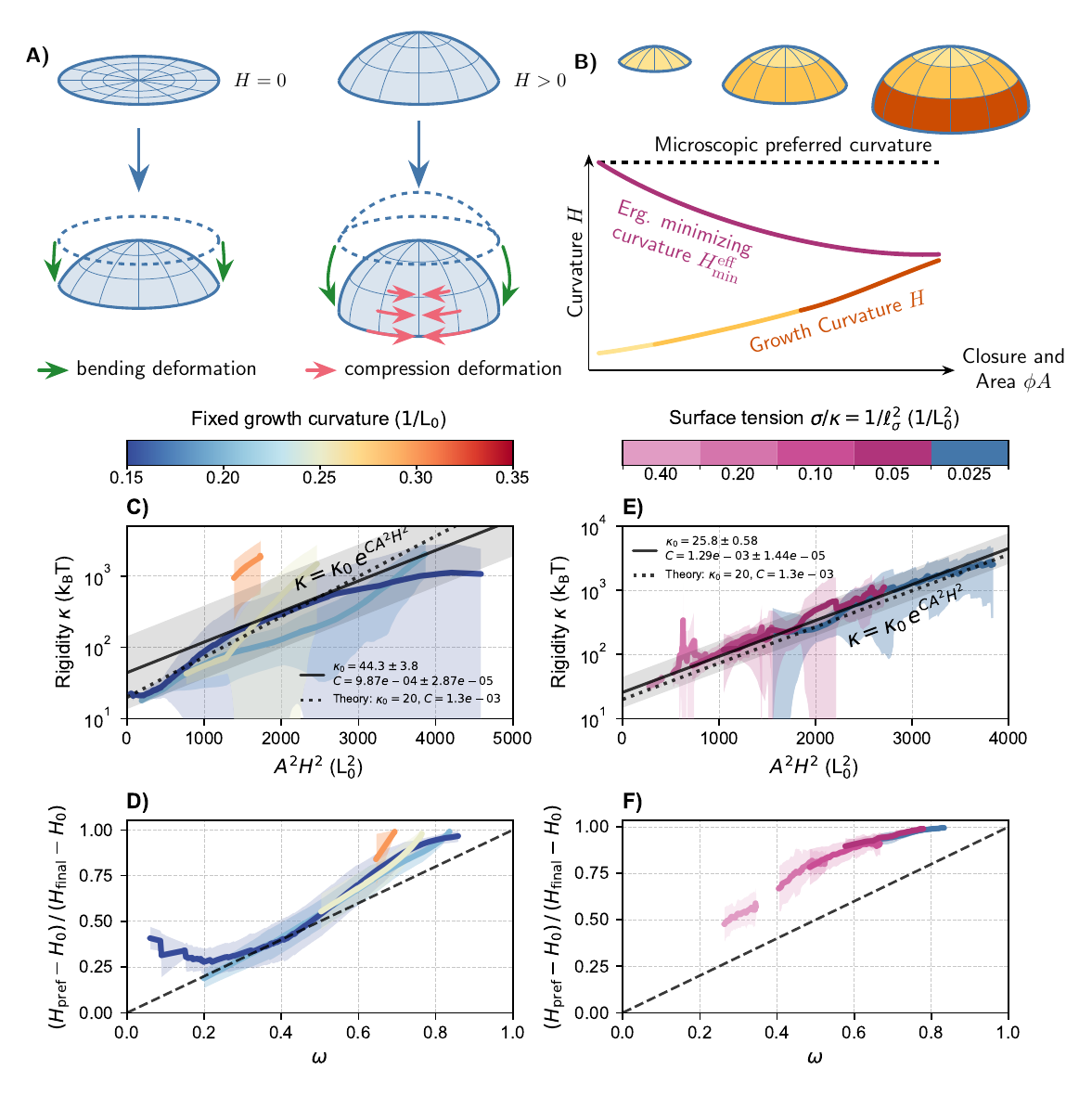}

	\captionof{figure}{\textbf{A non-Euclidean theory for geometry-induced stiffening and curvature memory mechanism.}
		(\textbf{A}) Curvature perturbations of a flat connected patch mainly
        excite angular bending modes, whereas perturbations of a
        curved cap also require compression or stretching. This geometric
        constraint produces the bending-to-stretching crossover. (\textbf{B})
        During growth, newly incorporated rings are added at the current
        membrane curvature and thereby store a local reference geometry; the
        mature coat therefore remembers a weighted average of its growth
        curvatures. (\textbf{C}, \textbf{D}) Theory and simulation at fixed
        growth curvature, with colors as in Fig.~\ref{fig:FixedCurvature}, $1\sigma$ error band as shaded region.
        The transformed effective coat bending rigidity collapses onto the
        predicted linear trend in (\textbf{C}), and the transformed
        energy-minimizing curvature follows the predicted weighting variable
        in (\textbf{D}). Dashed lines show the theoretical prediction and
        solid lines show simulation fits. (\textbf{E}, \textbf{F}) The same
        analysis for growth on a deformable membrane, with colors indicating
        the bending length as in Fig.~\ref{fig:VariableCurvature} and the $1\sigma$ error band shaded. The
        stiffness collapse in (\textbf{E}) matches the geometric prediction.
        In (\textbf{F}), using the final curvature as a proxy for the full
        curvature history shifts the data relative to the unshifted reference
        line, as expected for history-weighted curvature memory.}
	\label{fig:EffectiveMechanism} 
\end{figure*}

\subsection*{Curved geometry drives stiffening through a bending-to-stretching transition}

The fate separation above suggests a mechanical threshold. 
Once the lattice
is connected enough to transmit stress over the coat, curved spherical
geometry removes deformation modes that are available to a flat or poorly
connected lattice. Loose plaques remain outside this continuum description,
whereas connected curved coats can store geometric frustration and convert it
into an effective bending stiffness.

The mechanism is illustrated in Fig.~\ref{fig:EffectiveMechanism}A. A nearly
flat connected patch can change curvature mainly through angular
reorientation of clathrin legs, governed by the soft shear scale
$\mu\sim k_{\rm angle}\sim20\,k_{\rm B}T$. A curved cap is different: changing its
curvature changes the metric of a finite spherical patch, and a connected
lattice cannot relax this metric change everywhere by shear alone. A residual
area strain remains, forcing local compression or stretching and thereby
activating the much larger bulk scale
$K\sim k_{\rm stretch}\sim800\,k_{\rm B}T$. This bending-to-stretching crossover follows the
logic of incompatible elastic sheets and non-Euclidean elasticity
\cite{efratiElasticTheoryUnconstrained2009,
sharonMechanicsNonEuclideanPlates2010,
meiriCumulativeGeometricFrustration2021,kleinShapingElasticSheets2007}; the
full derivation of a corresponding theory is given in Supplementary Notes 1.2 and 1.3, the connection to non-Euclidian elasticity is made in 1.4, and the main results are explained in the following.

The companion mechanism, curvature memory, is illustrated in
Fig.~\ref{fig:EffectiveMechanism}B. Each newly incorporated ring of clathrin is
added in a mechanically relaxed state at the membrane curvature present at that
moment. As the coat subsequently grows and bends the membrane, older material
therefore carries a reference geometry written at earlier, usually lower,
curvature, whereas newer material carries the reference geometry of later,
more curved states. The energy-minimizing curvature of the assembled coat is
thus not only a microscopic property of individual triskelia; it is an
accumulated, weighted record of the curvatures at which connected material was
added.

For a small spherical patch, a relative curvature strain
$e=-\Delta H/H$ leaves a residual dilation $\Xi\simeq -4(\mu/K)e\phi$.
This result makes the geometric origin of the effect explicit. In the flat limit, $\phi\rightarrow0$, a curvature
perturbation does not excite the bulk modulus to leading order. At finite
closure, the residual dilation grows linearly with $\phi$, and the associated
stretching energy penalty, $K\,\Xi^2/2$, grows quadratically. We therefore evaluate the
coat energy under a curvature perturbation as a Helfrich-like bending response
augmented by the additional dilation-field contribution. Both contributions are quadratic, leading to new, effective parameters for the bending energy:
$\kappa_{\rm cl}^{\rm eff}$ and $H_{\rm min}^{\rm eff}$.

For coats assembled at a fixed growth curvature $H_{\rm growth}$, the theory
reduces to a single geometric weight $w$,
\begin{equation}
    w=C A^2H_{\rm growth}^2=4\pi C A\phi\ ,
\end{equation}
with $C=\mu/(2\pi^2K L_0^2)$. With the microscopic parameters used here,
$C\simeq1.3\times10^{-3}L_0^{-2}$. The predicted effective mechanics are
\begin{equation}
    \begin{split}
    \kappa_{\rm cl}^{\rm eff}&=\kappa_{\rm bend}^{\rm micro}\exp(w),\\
    H_{\rm min}^{\rm eff}
    &=(1-\omega)H_{\rm min}^{\rm micro}+\omega H_{\rm growth},
    \end{split}
\end{equation}
where $\omega=w/(1+w)$ interpolates
between the short and long term limits. $\kappa_{\rm bend}^{\rm micro}\sim\mu L_0^2\sim k_{\rm angle}$ is the
microscopic bending scale set by angular deformation of clathrin legs. Thus
the same geometric weight stiffens the coat and shifts its effective
energy-minimizing curvature from the microscopic value
$H_{\rm min}^{\rm micro}$ toward the curvature at which the material was
assembled. The curvature-memory law is linear in the weight: for $\omega=0$
the coat retains $H_{\rm min}^{\rm micro}$, whereas for $\omega\rightarrow1$ it
approaches $H_{\rm growth}$. In the fixed-curvature simulations
(Fig.~\ref{fig:EffectiveMechanism}C), the fitted stiffness trend,
$C=(9.87\pm0.29)\times10^{-4}L_0^{-2}$, is close to the prefactor predicted
from the independently chosen microscopic parameters $\mu$ and $K$. The
nontrivial transformation by $A^2H_{\rm growth}^2$ collapses data taken at
different imposed curvatures onto a common trend. Deviations are visible mainly
for the highest imposed curvatures, where a coat rapidly spans a large fraction
of its underlying sphere and the small-patch and quasi-static assumptions are
most strongly strained. The transformed energy-minimizing curvatures collapse more
tightly onto the predicted weighting variable $\omega =w/(1+w)$ 
(Fig. \ref{fig:EffectiveMechanism}D). With this theory, we have identified
the physical origin of the sigmoidal growth laws observed in experiments
and simulations.

During growth on a deformable membrane, material is incorporated at a sequence
of curvatures rather than at a single $H_{\rm growth}$. We therefore extend the fixed-curvature
result phenomenologically by assigning each incorporated material increment an individual weight $w_i$ and an incorporation curvature $H_i$. The relevant
accumulated quantities are the total stiffening weight and the normalized
curvature memory,
\begin{equation}
    \mathcal{W}=\sum_i w_i,\qquad
    \mathcal{H}=\frac{\sum_i w_iH_i}{\sum_i w_i}.
\end{equation}
The effective mechanics then become
\begin{equation}
    \begin{split}
    \kappa_{\rm cl}^{\rm eff}&=\kappa_{\rm bend}^{\rm micro}\exp(\mathcal{W}),\\
    H_{\rm min}^{\rm eff}
    &=(1-\omega)H_{\rm min}^{\rm micro}+\omega\mathcal{H},
    \end{split}
\end{equation}
with $\omega=\mathcal{W}/(1+\mathcal{W})$.
Curvature memory therefore does not mean that the coat simply remembers its
final shape. It remembers the weighted average of the curvatures at which its
connected material was added.

The deformable-membrane simulations provide the strongest test of this
framework. The same theoretical prefactor
computed from $\mu$ and $K$,
$C\simeq1.3\times10^{-3}L_0^{-2}$, matches the simulation fit,
$C=(1.29\pm0.014)\times10^{-3}L_0^{-2}$ (Fig. \ref{fig:EffectiveMechanism}E). The preferred-curvature collapse is
more phenomenological because the full history variable $\mathcal{H}$ is not
directly known from the main-figure readout (Fig. \ref{fig:EffectiveMechanism}F). Using the final curvature as a
proxy gives the expected offset from the unit-slope reference line while still
preserving the predicted trend. Because the final growth curvature is larger
than most curvatures sampled during assembly, this rescaling overestimates the
history-averaged memory $\mathcal{H}$ and should place the collapsed data above
the origin line. The fact that the points nevertheless align on a common line
supports the history-weighted memory picture. Together, the collapse of
$\kappa_{\rm cl}^{\rm eff}$ and the history-dependent shift of
$H_{\rm min}^{\rm eff}$ show that coat maturation creates the mechanical
inputs required for the area--curvature energy landscape introduced next.
That also means a shift in the description: the coat
is no longer characterized only by its instantaneous area and curvature, but by
history-dependent effective parameters that reshape the invagination energy
landscape.

\begin{figure*}
	\centering
	\includegraphics[width=1.0\textwidth, angle=0]{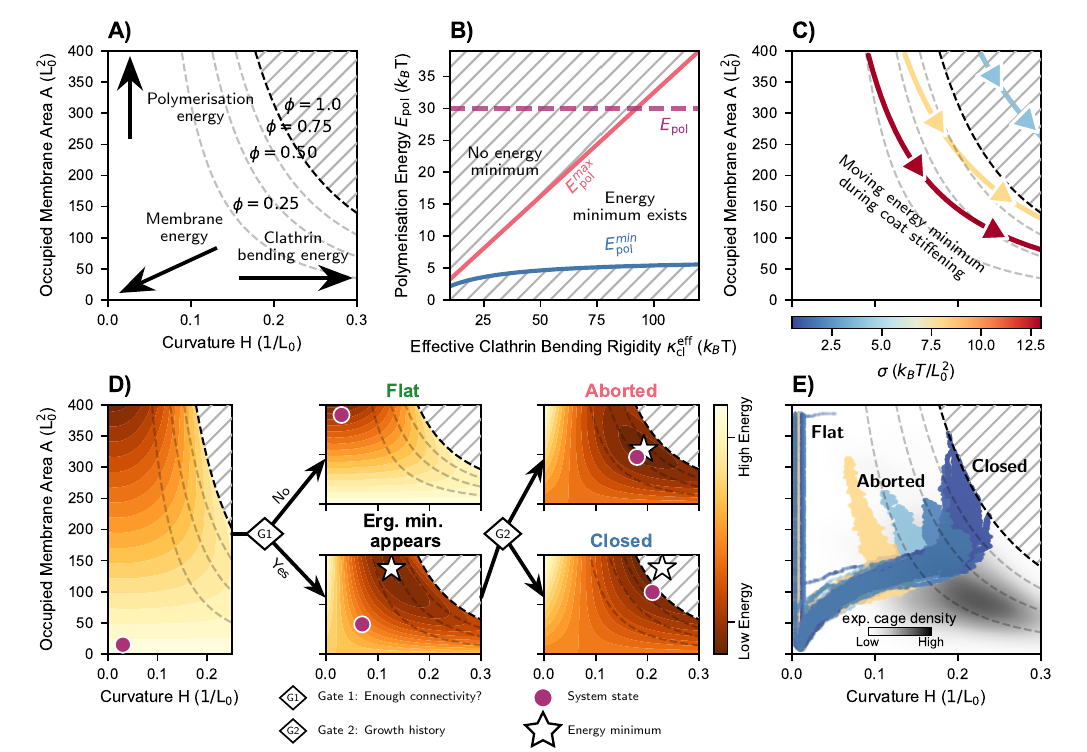}

	\captionof{figure}{
    \textbf{Coat stiffening moves the energy minimum and thereby decides fate.}
  (\textbf{A}) The shape of the energy landscape is controlled by three contributions, the polymerization 
  energy, the membrane bending and surface tension energy, and the clathrin bending energy. The three
  contributions push the system into three different directions in the curvature-area plane. 
  (\textbf{B}) A finite energy minimum exists only when the
  polymerization energy per triskelion lies within a stiffness-dependent
  window. As $\kappa_{\rm cl}^{\rm eff}$ increases during growth, a fixed
  polymerization energy can enter this window and create an attractor. (\textbf{C}) For different
  surface tensions $\sigma$, the trajectory of
  the energy minimum is shifted and crosses the closure line $\phi=1$ at different points. (\textbf{D}) Energy landscapes at two sequential fate gates. Initially, polymerization energy dominates and the system reduces its energy by expanding area at low curvature. At gate G1 (connectivity), insufficient coat connectivity confines the system to the Flat state; once sufficient connectivity is established, the effective stiffness rises into the window that supports an energy minimum (B), and a minimum appears at finite curvature, redirecting the system toward higher $H$. The minimum then migrates through the landscape according to surface tension and growth history (C). At gate G2 (growth history), the final fate depends on whether the minimum is reached within the accessible region (Aborted), or crosses the closure boundary before (Closed).
  (\textbf{E}) Simulated trajectories colored by surface tension $\sigma$,
  with experimentally observed clathrin pits overlaid in gray
  \cite{tagiltsevNanodissectedElasticallyLoaded2021}.}
	\label{fig:EnergyLandscape} 
\end{figure*}

\subsection*{Coat connectivity and stiffening control the energy minimum and thereby fate}

The stiffening and curvature memory derived above allow us to turn the
simulation trajectories into a coarse-grained mechanical picture. We describe
an assembling pit by its coated membrane area $A$ and curvature $H$, and ask
how its effective energy landscape $E(A,H)$ changes as the coat matures
(Fig.~\ref{fig:EnergyLandscape}). This landscape is not the microscopic KMC
Hamiltonian. Instead, it summarizes the many lattice configurations accessible
to a coat of a given size and curvature by two history-dependent material
properties: the effective coat bending rigidity $\kappa_{\rm cl}^{\rm eff}$ and the
effective energy-minimizing curvature $H_{\rm min}^{\rm eff}$. In this way,
the discrete assembly history derived above becomes a moving continuum
description of the invaginating pit. A detailed derivation with explicit calculations is found in Supplementary Note 1.5.

Fig.~\ref{fig:EnergyLandscape}A summarizes the three physical contributions, whose exact form can be found in the Methods.
Clathrin polymerization lowers the energy as more triskelia are incorporated
and therefore drives growth in area. Membrane bending rigidity $\kappa_{\rm M}$ and surface tension $\sigma$ penalize curved and excess membrane area and therefore oppose deep
invagination~\cite{tagiltsevNanodissectedElasticallyLoaded2021,
saleemBalanceMembraneElasticity2015}. Clathrin elasticity penalizes deviations
from the coat's current effective energy-minimizing curvature $H_{\rm min}^{\rm eff}$, with a strength set by
$\kappa_{\rm cl}^{\rm eff}$, and is the contribution that changes in strength. The landscape therefore shows how growth,
membrane resistance, and coat elasticity compete in the same area--curvature
state plane. This connects our model to previous energy-landscape descriptions
of endocytic invagination~\cite{tagiltsevNanodissectedElasticallyLoaded2021,
saleemBalanceMembraneElasticity2015,freyCompetingPathwaysInvagination2020}.

As the coat matures, the polymerization energy $\Delta E_{\rm pol}$ is treated
as a fixed microscopic input, while $\kappa_{\rm cl}^{\rm eff}$ and
$H_{\rm min}^{\rm eff}$ evolve with growth history. Early on, energy can be lowered mainly by increasing coated area at low curvature through the polymerization term. Once
connectivity and curved geometry generate sufficient coat stiffness, the
clathrin elastic term becomes strong enough to reshape the landscape and pull
the coat toward higher curvature (Fig. \ref{fig:EnergyLandscape}A). Coat stiffening therefore does not merely make
an already curved coat harder to deform; it shifts the mechanical state that
the assembling pit is trying to reach.

At each point along an evolution trajectory, the current values of
$\kappa_{\rm cl}^{\rm eff}$ and $H_{\rm min}^{\rm eff}$ define an
instantaneous landscape. Reshaping of this landscape can be expressed as the appearance and movement
of an energy minimum at finite area and curvature (Fig. \ref{fig:EnergyLandscape}B, C). With $A_{\rm cl}$ denoting
the mean coated area per incorporated triskelion, a finite basin exists only
when the polymerization energy lies within a stiffness-dependent window,
\begin{equation}
    \Delta E_{\rm pol}^{\rm min}
    < \Delta E_{\rm pol}
    < \Delta E_{\rm pol}^{\rm max},
\end{equation}
with
\begin{equation}
    \Delta E_{\rm pol}^{\rm max}
    = A_{\rm cl}\left(H_{\rm min}^{\rm eff}\right)^2
    \kappa_{\rm cl}^{\rm eff},
    \qquad
    \Delta E_{\rm pol}^{\rm min}
    = \Delta E_{\rm pol}^{\rm max}
    \frac{\kappa_{\rm M}}{\kappa_{\rm M}+\kappa_{\rm cl}^{\rm eff}}.
\end{equation}
The full derivation and the corresponding expressions for $A_{\rm crit}$ and
$H_{\rm crit}$, the coordinates of the energy minimum, are given in the Supplement and the Methods respectively. If polymerization is too weak, growth
is not favorable $(A,H\rightarrow0)$. If it is too strong, the system
can continue lowering its energy by adding area and no finite basin arrests
the trajectory; this is the landscape analogue of rapid flat plaque growth
when the clathrin term never becomes strong enough to bend the membrane.
Increasing $\kappa_{\rm cl}^{\rm eff}$ moves a fixed $\Delta E_{\rm pol}$ into
the existence window (Fig. \ref{fig:EnergyLandscape}B), so a (moving) minimum appears. At fixed ratio of surface tension $\sigma$ to membrane rigidity $\kappa_{\rm M}$, the trajectory of the minimum starts at high area and low curvature, and moves with increasing effective bending rigidity $\kappa_{\rm cl}^{\rm eff}$ towards lower areas and higher curvatures. Increasing $\sigma/\kappa_{\rm M}$ shifts the position of the minimum trajectory to the origin (Fig.~\ref{fig:EnergyLandscape}C).

The three fates emerge from a branched assembly trajectory governed by two sequential fate gates (Fig.~\ref{fig:EnergyLandscape}D). At early times, polymerization drives area growth, and the energy landscape favors a large, flat coat. The first gate is whether the coat develops sufficient connectivity to enter the stretching-dominated regime: coats that fail to do so remain floppy, continue growing flat, and arrest as persistent plaques. Coats that cross this threshold stiffen, and an energy minimum at finite area and curvature appears in the landscape (Fig.~\ref{fig:EnergyLandscape}B), pulling the system toward an invaginated state. The second gate is determined by a race between the system and its own energy minimum, both of which move through the landscape as the coat stiffens. If the minimum escapes beyond the closure boundary ($\phi=1$) before the system reaches it, closure is forced; if the system catches its minimum first, it arrests as an aborted pit.

Borderline cases are expected because this landscape is an instantaneous
coarse-grained description, not a sharp microscopic phase boundary. As a
trajectory approaches its moving minimum, the driving force decreases
continuously and closure can slow strongly before motion stops. In biological
terms, these coats may still have a weak residual drive toward closure, but
not enough to close within the finite lifetime available to an endocytic site.
This connects the slow or stalled simulated trajectories to abortive cellular
events without treating the simulated observation window as the underlying
physical mechanism.

The simulated trajectories in Fig.~\ref{fig:EnergyLandscape}E follow this
logic. Lower ratio of membrane surface tension to membrane bending rigidity $\sigma/\kappa_{\rm M}$, or larger bending length $\ell_\sigma$, shifts the moving
minimum so that coats reach higher curvature and more often close; higher
tension keeps trajectories closer to the low-curvature region and favors
stalling. The experimentally observed pit geometries from \cite{tagiltsevNanodissectedElasticallyLoaded2021} overlaid in gray occupy
the same order-of-magnitude area--curvature range as the simulated stalled and
closing trajectories. They are found at lower areas, because the AFM experiments mainly 
observe partially closed pits, but agree well with the curvature predicted by our simulations.


\begin{figure*}
	\centering
	\includegraphics[width=1.0\textwidth, angle=0]{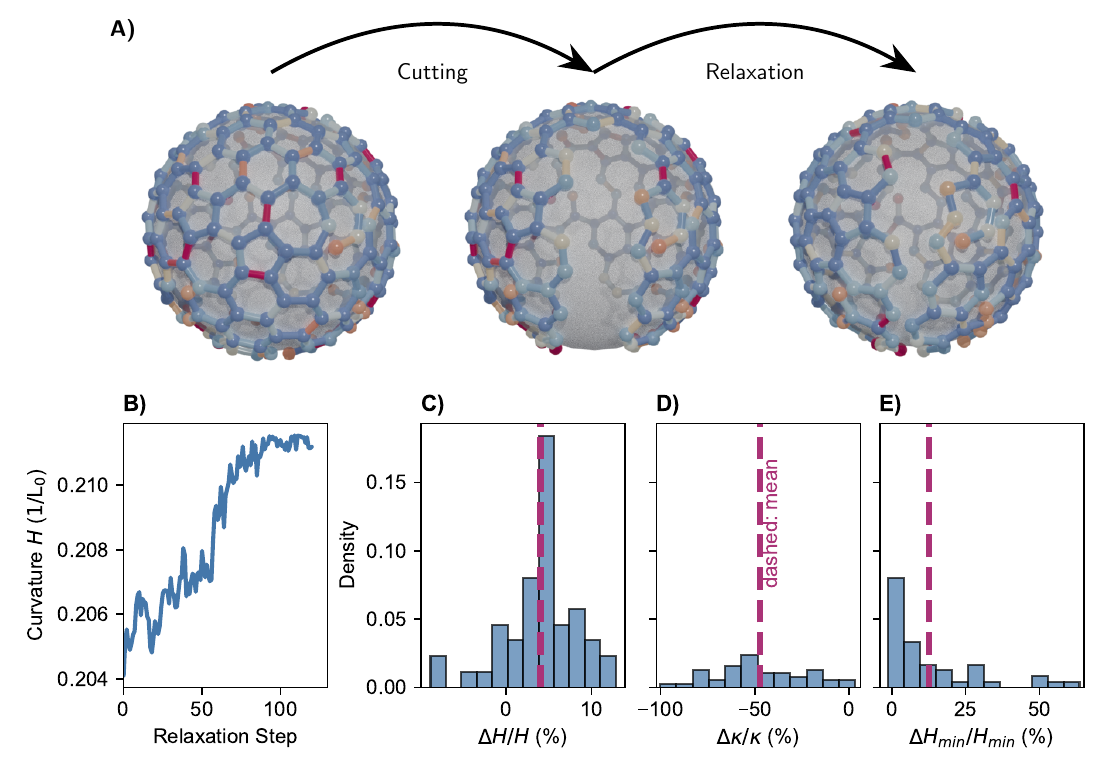} 
	
	\caption{\textbf{Lattice cutting reduces stiffness and releases curvature memory.}
  (\textbf{A}) In silico cutting protocol: an assembled coat is selected, a
  radial geodesic is defined, bonds and nodes along the cut are removed, and
  the remaining connected lattice is allowed to relax. Bond and node colors encode the
  local energetic load on a logarithmic scale (blue is low, red is high). (\textbf{B}) 
  Example trajectory
  showing curvature relaxation after cutting. (\textbf{C}) Distribution of
  relative curvature changes, with a positive mean. (\textbf{D}) Distribution
  of relative effective bending-rigidity changes, with a negative mean.
  (\textbf{E}) Distribution of relative preferred-curvature changes, with a
  positive mean.}
	\label{fig:CuttingExperiments} 
\end{figure*}

\subsection*{In silico cutting validates growth-history and stiffening framework}

A direct test of prestress and curvature memory is to perturb an assembled, connected coat.
Tagiltsev et al.~\cite{tagiltsevNanodissectedElasticallyLoaded2021} used
HS-AFM nanodissection to cut plasma-membrane-associated clathrin coats and
observed relaxation toward increased curvature. We reproduce this logic in
silico by selecting coats whose rapid shape evolution has stalled, severing
bonds along a radial geodesic from the pole, and then allowing the remaining
connected lattice to relax (Fig.~\ref{fig:CuttingExperiments}A). In a typical
trajectory, the coat relaxes toward a higher curvature after the cut
(Fig.~\ref{fig:CuttingExperiments}B), and the same tendency is observed across
the ensemble (Fig.~\ref{fig:CuttingExperiments}C).

To connect the geometric response to the mechanical framework, we infer
$\kappa_{\rm cl}^{\rm eff}$ and $H_{\rm min}^{\rm eff}$ before and after
cutting. The cut consistently reduces the inferred bending rigidity, with an
average reduction of roughly 50\% (Fig.~\ref{fig:CuttingExperiments}D). At the
same time, the inferred energy-minimizing curvature increases
(Fig.~\ref{fig:CuttingExperiments}E). Thus cutting does not simply release a
stored stress or soften the coat in isolation. It changes both state variables
of the growth-history framework: the remaining coat becomes easier to deform,
but it also relaxes toward a higher effective energy-minimizing curvature.

The ring-additive theory predicts these two effects. We represent a cut by
reducing the weight of the affected material increment, from $w_i$ to $c_iw_i$, where
$c_i$ is the retained mechanical fraction of increment $i$ after cutting,
\begin{equation}
    w_i\rightarrow c_iw_i,
    \qquad 0\leq c_i\leq1.
\end{equation}
The removed stiffening weight $\delta W$ and the curvature memory carried by the removed
material $H_{\rm cut}$, with $H_i$ again denoting the incorporation curvature of increment
$i$, are
\begin{equation}
    \delta W=\sum_i(1-c_i)w_i,
    \qquad
    H_{\rm cut}=
    \frac{\sum_i(1-c_i)w_iH_i}{\delta W}.
\end{equation}
Because the post-cut coat has less accumulated stiffening weight,
\begin{equation}
    \Delta\ln\kappa_{\rm cl}^{\rm eff}\simeq-\delta W<0,
\end{equation}
so cutting lowers the effective coat bending rigidity.
The shift in energy-minimizing curvature is
\begin{equation}
    \Delta H_{\rm min}^{\rm eff}
    \simeq
    \frac{\delta W}{1+\mathcal{W}}
    \left(H_{\rm min}^{\rm eff}-H_{\rm cut}\right).
\end{equation}
Its sign depends on the curvature memory removed by the cut. If the severed
material was incorporated at lower curvature than the intact coat, which is
usually the case in the observed assemblies, then
$H_{\rm cut}<H_{\rm min}^{\rm eff}$ and cutting increases the effective
energy-minimizing curvature. A more detailed derivation is found in Supplementary Note 1.6.

This is the regime sampled by the in silico cuts. They reduce the
stress-transmitting weight of the connected lattice while preferentially
removing low-curvature memory from earlier growth. The experimental
nanodissection response, the first-order theory, and the simulated perturbation
therefore agree on the same mechanical interpretation: assembled clathrin
coats store growth history in their connected lattice, and cutting reveals
this memory by making the coat softer while shifting its relaxed curvature
upward.

\subsection*{Discussion}

Clathrin assembly has no single morphological endpoint: the same basic coat
architecture can remain flat, stall after partial invagination, or close into
a vesicle. A physical theory of CME must therefore explain how one assembling lattice
accesses several fates without assigning an independent mechanism to each.
Our results identify maturation of coat mechanics as a relevant intermediate layer. During
growth, the coat not only accumulates clathrin mass; it changes its capacity
to transmit stress, stiffens through geometry, and shifts the curvature it
mechanically prefers. Pathway variability can thus arise from the evolving
mechanical state of the coat itself. 
More generally, the discrete effects of self-assembly influence the effective
parameters of any continuum description, which can also be expected in
other systems, especially if the number of individual agents are small, 
and finite size effects therefore more important.

Our simulations and theory reevaluate what should be considered 
input and output of CME. In many continuum descriptions of CME, coat stiffness 
and preferred curvature are prescribed, and the membrane shape
is then calculated from them. Only recently 
has it been suggested that a coat rigidity might be a dynamical
variable \cite{freyCoatStiffeningCan2024a}, but the underlying microscopic
mechanisms were not addressed. Our simulation framework now demonstrates
how coat rigidity and preferred curvature are dynamically generated
by the interplay of many discrete microscopic processes.
It connects constant-curvature models, constant-area models, plaque-derived
budding scenarios, and polymerization--tension landscapes~\cite{
avinoamEndocyticSitesMature2015,mundClathrinCoatsPartially2023,
lampeClathrinCoatedPits2016,saleemBalanceMembraneElasticity2015,
tagiltsevNanodissectedElasticallyLoaded2021,
freyCompetingPathwaysInvagination2020}: each can be viewed as a limiting
description of a lattice whose effective continuum parameters are themselves
assembly-generated.

The effective coat bending rigidity emerges from curved spherical geometry.
Mature coats are much stiffer than estimates based on individual triskelion
flexibility~\cite{jinRigidityTriskelionArms2000,
nossalEnergeticsClathrinBasket2001,jinMeasuringElasticityClathrinCoated2006,
tagiltsevNanodissectedElasticallyLoaded2021}. In the model, this gap does not
require a hidden stiff molecular state. Weakly curved or weakly connected
coats can change curvature mainly by angular rearrangement; curved
coats necessarily generate residual dilation and compression. The response
therefore shifts from bending-dominated to stretching-dominated without
changing the microscopic rules.

The effective energy-minimizing curvature is generated at the same time. Each newly
incorporated triskelion contributes a local reference geometry set by the
curvature at incorporation. The energy-minimizing curvature of the mature coat is
therefore encoded in the current mechanical state of the lattice rather than
fixed only by molecular architecture. This places clathrin assembly in the
broader mechanics of incompatible and growing elastic bodies~\cite{
efratiElasticTheoryUnconstrained2009,sharonMechanicsNonEuclideanPlates2010,
meiriCumulativeGeometricFrustration2021,kleinShapingElasticSheets2007} and
offers a physical explanation for why inferred preferred curvatures can vary
across cell types, preparations, or mechanical environments.

Fate selection then requires two mechanical steps. First, the coat must
transmit stress over the assembly. In the simulation this appears as a
connectivity gate, because clathrin--clathrin bonds are the explicit route of
stress transmission. Biologically, the same mechanical state could also arise
from loose contacts, weak or transient bonds, hidden vacancies, adaptor state,
adhesion-associated constraints, or other local regulation. The general
prediction is therefore that flat coats fail to transmit coherent bending
stresses, not that they must share a universal connectivity threshold.

Second, once stress transmission is possible, fate is set by the moving
energy landscape. Geometry-induced stiffening strengthens the
clathrin elastic term, curvature memory shifts the energy-minimizing curvature, and
membrane mechanics determine whether the evolving minimum stalls inside the
accessible region or crosses the closure boundary. The bending length
$\ell_\sigma=\sqrt{\kappa_{\rm M}/\sigma}$ condenses membrane bending rigidity and
surface tension into the scale controlling this competition. Molecular
regulators enter this framework through membrane load, polymerization gain,
lattice coupling, curvature preference, and kinetic rates rather than through
a separate theoretical term for every protein.

This view also separates curvature generation from topological completion. A
closed trivalent cage must contain twelve excess pentagons~\cite{
jinTopologicalMechanismsInvolved1993,morrisCryoEMMultipleCage2019}, but this
does not imply that pentagons initiate bending. In the simulations, curvature
can begin before visible non-hexagonal polygons are counted, whereas flat
lattices can contain defects without curving. Defects are therefore best
understood as topological elements that accommodate and stabilize a curved
state generated by connected coat mechanics, consistent with observations of
flat, defect-containing lattices~\cite{
sochackiStructureSpontaneousCurvature2021}.

The most direct tests are mechanical. Curvature-controlled assembly on beads
or curved membrane supports could test whether stiffness increases with
coated area and growth curvature. AFM indentation and nanodissection can test
whether cutting a connected coat reduces its effective stiffness and shifts
its relaxed curvature~\cite{tagiltsevNanodissectedElasticallyLoaded2021}.
Structural methods such as cryo-ET and platinum-replica EM remain essential,
but the target is mechanical coupling rather than a universal connectivity
value: contact density, vacancies, bond geometry, lattice disorder, and
adaptor organization are possible readouts that could be explicitly linked to stress transmission. 
Controlled tension perturbations should shift stalled and closed outcomes through
$\ell_\sigma$~\cite{saleemBalanceMembraneElasticity2015,
bucherClathrinadaptorRatioMembrane2018,hassingerMembraneTensionKey2016}.

A further theoretical extension would be to relax the spherical-cap constraint.
Our simulations describe invagination by a single mean curvature and area,
which isolates the role of assembly history but excludes non-spherical
deformation modes. Allowing anisotropic deformable pits could further explain
how local the curvature memory is, and to what extent the system naturally 
prefers symmetrical, and hence spherical configurations.

The model presented here demonstrates how the microscopic mechanics of single 
triskelia alone can lead to macroscopically different cage fates.
Cargo, adaptors, epsin, BAR-domain proteins,  actin, dynamin, lipid composition, 
adhesion, turnover, and scission can and will also 
reshape the effective parameters that control coat maturation and fate.~\cite{fordCurvatureClathrincoatedPits2002, peterBARDomainsSensors2004, zimmerbergHowProteinsProduce2006} 
In our treatment, the environment shapes CME through surface
tension, but in future extensions, one could include e.g. the
effects of adhesion to differently sized cargo or the role
of actin networks pushing and pulling on the growing pit.

Our unifying framework demonstrates that physical processes
determine the main fate decisions. Because biochemical regulation can 
shift the balance at these gates, physics and biochemistry will
go hand in hand when determining CME fate decisions in different cell types
and in different environments. CME should therefore be viewed not as a set of separate 
pathways for plaques, stalled pits, and vesicles, but as a regulated self-assembly 
process whose biological diversity is organized by mechanics that is
generated by the growth process itself, leading to many feedback loops
and high fate variability.

\newpage
\clearpage
\subsection*{Methods}

\begin{table*}[!htbp]
    \centering
    \caption{\textbf{Experimental values and literature benchmarks for clathrin-mediated endocytosis.}
    Values are taken from the literature or derived from model geometry; entries used as direct simulation
    inputs are indicated. In the Value column each reported number is immediately followed by its source
    in square brackets; all sources are collected again in the Reference column.
    $L_0=18.4\,\mathrm{nm}$ is used throughout for unit conversion.}
    \label{tab:key parameters}
    \scriptsize
    \renewcommand{\arraystretch}{1.25}
    
    \begin{tabular}{p{0.11\textwidth}|p{0.27\textwidth}|p{0.37\textwidth}|p{0.14\textwidth}}
        \textbf{Parameter} & \textbf{Description} & \textbf{Value} & \textbf{References}\\
        \hline
        \hline
        $L_0$ & Clathrin-clathrin lattice bond length (hub-to-hub); normalization length throughout.&
        $\sim18.4\pm2.0\,\mathrm{nm}$
        &
\cite{jinTopologicalMechanismsInvolved1993,vigersThreedimensionalStructureClathrin1986,crowtherAssemblyPackingClathrin1981}\\
\hline
        $\sigma$ & Membrane tension. &
        $\sim 1-30~\mu\mathrm{N}\,\mathrm{m}^{-1}\approx0.25\times 10^{-3}-7.5\times 10^{-3}\kbt\,\mathrm{nm}^{-2}\approx 0.1-2.5~\kbt\,\mathrm{L}_0^{-2}$
        &
        \cite{daiMechanicalPropertiesNeuronal1995, raucherCellSpreadingLamellipodial2000}\\
        \hline
        $\kappa_\mathrm{M}$ & Membrane bending rigidity. &
        ${\sim}10$--$25\,\kbt$, in simulations $\kappa_{\rm M}=20~\kbt$
         &
\cite{rawiczEffectChainLength2000,evansEntropydrivenTensionBending1990}\\
        \hline
        $\kappa_\mathrm{cl}^{\rm mature}$ &  Bending rigidity of mature clathrin coats/CCVs. &
        ${\sim}100$--$400\,\kbt$
&
        \cite{jinMeasuringElasticityClathrinCoated2006, tagiltsevNanodissectedElasticallyLoaded2021}\\
        \hline
        $EI_\mathrm{arm}$ & Flexural rigidity of one clathrin leg. &
        ${\sim}35\,\kbt\,\mathrm{nm}$ &
        \cite{jinRigidityTriskelionArms2000}\\
        \hline
        $k_\mathrm{bend}$ & Angular leg bending constant, related to 2D shear modulus
        $\mu=k_\mathrm{bend}/L_0^2$, calibrated to $EI_\mathrm{arm}$. &
        $k_\mathrm{bend}=20~\kbt$
         &
\cite{jinRigidityTriskelionArms2000}\\
        \hline
        $k_\mathrm{stretch}$ & Leg stretching constant, related to 2D bulk modulus
        $K=k_\mathrm{stretch}/L_0^2$, calibrated from $L_0$ fluctuations. &
        $k_\mathrm{stretch}=800\,\kbt$&
        \cite{jinRigidityTriskelionArms2000}\\
        
        \hline
        $A_\mathrm{cl}$ & Area per triskelion in the coat. &
        ${\sim}433\,\mathrm{nm}^2=1.28\,L_0^2$, in simulations $A_{\rm cl}=1.3 L_0^2$
        &
        \cite{vigersThreedimensionalStructureClathrin1986,crowtherAssemblyPackingClathrin1981}\\
        \hline
        $\Delta E_\mathrm{pol}$ & Polymerization/binding energy gain per triskelion upon coat incorporation. &
        ${\sim}5$--$30\,\kbt$, in simulations $\Delta E_{\rm pol}=1~\kbt$ 
        &
        \cite{denotterGenerationCurvedClathrin2011,nossalEnergeticsClathrinBasket2001,
              saleemBalanceMembraneElasticity2015}\\
        \hline
        $N_\mathrm{CC}$ & Number of triskelia in a typical in-vitro clathrin cage. &
        ${\sim}40$
      
          &
        \cite{vigersThreedimensionalStructureClathrin1986}\\
        \hline
        $R_\mathrm{CC}$, $H_\mathrm{CC}$ & Radius and mean curvature of in-vitro clathrin cages, microscopic energy-minimizing parameter $H_{\rm min}^{\rm micro}$. &
        $R_\mathrm{CC}{\approx}30\,\mathrm{nm}=1.63\,L_0$;
        $H_\mathrm{CC}{\approx}32\,\mu\mathrm{m}^{-1}=0.61\,L_0^{-1}$
          &
        \cite{vigersThreedimensionalStructureClathrin1986}\\
        \hline
        $N_\mathrm{CCV}$ & Number of triskelia in a clathrin-coated vesicle. &
        ${\sim}60$--$140$
         &
        \cite{heuserDeepetchViewsClathrin1985}\\
        \hline
        $R_\mathrm{CCV}$, $H_\mathrm{CCV}$ & Radius and mean curvature of cellular CCVs/CCPs. &
        $R_{\rm CCV}{\sim}60$--$350\,\mathrm{nm}=3.24$--$19.02\,L_0$;
        $H_\mathrm{CCV}{\approx}16$--$2.8\,\mu\mathrm{m}^{-1}=0.30$--$0.05\,L_0^{-1}$ &
        \cite{crowtherStructureCoatedVesicles1976, heuserThreedimensionalVisualizationCoated1980, crowtherAssemblyPackingClathrin1981}\\
        \hline
        $\tau_\mathrm{CCV}$ & Mean lifetime of productive CCPs (ending as CCV). &
        ${\sim}30$--$120\,\mathrm{s}$
          &
        \cite{loerkeCargoDynaminRegulate2009, ehrlichEndocytosisRandomInitiation2004}\\
        \hline
        $\tau_\mathrm{abort}$ & Mean lifetime of aborted (non-productive) CCPs. &
        $\tau\lesssim20\,\mathrm{s}$
          &
        \cite{loerkeCargoDynaminRegulate2009, ehrlichEndocytosisRandomInitiation2004}\\

    \end{tabular}
\end{table*}

\subsubsection*{Individual clathrin bending rigidity} It is possible to infer an estimate for the bending rigidity analogue of a single clathrin. Following the work of Jin and Nossal~\cite{jinRigidityTriskelionArms2000}, the flexural rigidity of clathrin arms is $EI_{\rm arm}\approx 35~\kbt\,\mathrm{nm}$. A coat is formed from units whose arms bind together, so at least two, and at most four arms stick together to form a bond. The flexural rigidity of such a bond is\begin{equation}
    EI_{\rm bond}=gEI_{\rm arm}\,,
\end{equation}with $g\in[12,26]$ for four arms, depending on the geometric arrangement of the arms.

Imagine now a square of clathrin coat of length $L_0$. To bend such a square in one direction towards curvature $H=1/R$ requires bending two such rods, which costs\begin{equation}
    E_{\rm bend,rod}=\frac{EI_{\rm bond}L_0}{R^2}\,.
\end{equation}On the other hand, the classical Helfrich picture requires the energy\begin{equation}
    E_{\rm bend, Helfrich}=\frac{A\kappa}{R^2}
\end{equation}with $A=L_0^2$. Equating both yields\begin{equation}
    L_0\kappa=EI_{\rm bond}\Rightarrow \kappa_{\rm cl}^{\rm micro}=g\frac{EI_{\rm arm}}{L_0}\,.
\end{equation}With $EI_{\rm arm}/L_0\approx 1.9~\kbt$ this leaves\begin{equation}
    \kappa_{\rm cl}^{\rm micro}\sim 24-52~\kbt
\end{equation}where one can average this value over the involved legs, leading to $\kappa_{\rm cl}^{\rm micro}\sim 6-13~\kbt$. A more elaborate derivation can be found in Supplementary Note 1.1.

\subsubsection*{Microscopic Model}
Each clathrin triskelion is represented as a hub at position $\boldsymbol{x}$ that forms
connections ("bonds") to other hubs at positions $\textbf{y}$, with Euclidean 3D length
$L_y:=\sqrt{(\textbf{x}-\textbf{y})^2}$. Hubs also form next-nearest-neighbor bonds from $\textbf{y}$
to $\textbf{z}$, of length $L_z:=\sqrt{(\textbf{y}-\textbf{z})^2}$. Two bonds to direct neighbors,
$L_i$ and $L_j$, enclose a planar angle $\varphi_{ij}$, as does each next-nearest-neighbor bond with
its proximal leg, and every bond $i$ is tilted out of the local tangent plane by a pucker angle
$\chi_i$, see Fig.~\ref{fig:ModelExplanation}C. The microscopic clathrin energy is
\begin{equation}
    \begin{split}
        E_{\rm bonds}=\sum_i\biggl\lbrace& k_{\rm stretch}\left(\frac{L_i}{L_0}-1\right)^2\\
        +&k_{\rm bend}\biggl[(\chi_i-\chi_0)^2+(\varphi_i-\varphi_0)^2\biggr]\biggr\rbrace\,,
    \end{split}
\end{equation}
where the sum runs over formed bonds only. The equilibrium parameters are $L_0$, $\chi_0$ and
$\varphi_0$: geometry fixes $\varphi_0=2\pi/3$, $L_0$ is
clathrin's equilibrium leg length of $18.4~$nm, and $\chi_0$ encodes its microscopic preferred
curvature.

This energy is coupled to a membrane energy of Helfrich form~\cite{helfrichElasticPropertiesLipid1973},
\begin{equation}
    E_{\rm membrane}=A\kappa_{\rm M}H^2\,.
\end{equation}
Since the curvature is uniform in our geometry, no integration is required, and the membrane
preferred curvature is set to zero. The area $A$ is a weighted mean of $N_{\rm cl}A_{\rm cl}$, the
number of bound clathrins times the area per clathrin, and $A_{\rm enclosing}$, the smallest
spherical-cap area accommodating all bound clathrins. A polymerization energy $\Delta E_{\rm pol}$
is released whenever a previously unbound clathrin first attaches to the membrane.

\subsubsection*{Evolution algorithm}
The dynamics separate into a fast, adiabatic regime and a slow, time-evolving regime. In the fast
regime, hub diffusion and curvature adaptation are assumed to remain in quasi-equilibrium and are
sampled with a standard Metropolis Monte Carlo algorithm using sufficiently many steps at each
instant.

The slow regime drives the physical time evolution through a kinetic Monte Carlo
algorithm~\cite{sickafusRadiationEffectsSolids2007}, which assigns a transition rate $\gamma_i$ to
every possible event $i$. The Gillespie algorithm is a widely used special
case~\cite{gillespieExactStochasticSimulation1977}. We take each rate to follow an Arrhenius-like
law,
\begin{equation}
    \gamma_i=\gamma\exp\left(-\beta\Delta E_i\right),
\end{equation}
where $\Delta E_i$ is the energy change of process $i$, $\gamma$ is a base rate setting the overall
time scale, and $\beta$ is the inverse temperature; we work in units of $\kbt$, so $\beta=1$.
Clathrin binding events are the exception: because a newly attaching clathrin has a continuum of
possible attachment positions, sampling their transition probabilities requires a rejection
step~\cite{bortzNewAlgorithmMonte1975}, at the cost of reduced efficiency.

\subsubsection*{Implementation}
The model is implemented in \textsc{Python} using the \textsc{JAX} library, which provides just-in-time compilation of array programs.~\cite{jax2018github} A complete kinetic Monte Carlo trajectory, consisting of a fixed number of update steps, is expressed as a single compiled scan over the step index, so that event construction, event sampling, energy evaluation, and the acceptance step are fused into one traced computation rather than dispatched step by step through the Python interpreter. Compiling the entire evolution loop in this way, rather than its individual operations, is the principal source of the resulting speedup compared to classical \texttt{numpy}. The price of this approach is a more restrictive data layout, because traced computations require array shapes that are fixed at compile time: the lattice is held in node arrays of fixed size, in which unoccupied binding slots are marked by sentinel values, so that the coat grows by activating preallocated entries rather than by resizing arrays. The global simulation state and the node ensemble are registered as \textsc{JAX} pytrees whose definitions separate the dynamic arrays traced during compilation from the static, Python-side configuration, such as node counts and microscopic spring constants. This representation keeps the data structures close to the underlying physical objects while allowing the kinetic Monte Carlo loop to be compiled as a whole and dispatched without modification to either CPU or GPU hardware.

The first execution for a given array shape carries a one-time compilation cost, after which replicate ensembles run at the compiled speed. Each replicate is initialized from a recorded random seed, and all microscopic parameters are read from a tabulated parameter file, so that individual trajectories can be reproduced exactly. The simulations reported here were run on the bwHPC cluster; the 30 variable-curvature trajectories underlying Fig.~\ref{fig:VariableCurvature} required between two and four days on a single 64-core node. The complete simulation and analysis pipeline, together with a pinned \textsc{conda} environment and a minimal runnable example, is publicly available under the MIT License (see Data and materials availability), and described in more detail in Supplementary Note 2.

\subsubsection*{Area--curvature landscape\label{sec:area--curvature-landscape}}
For the coarse-grained invagination analysis, a clathrin-coated pit is described by its coated area
$A$ and the mean curvature $H=1/R$ of a spherical cap. The accessible state space is bounded by the
closure condition
\begin{equation}
    \phi=\frac{AH^2}{4\pi}\leq1,
    \qquad
    A\leq A_{\rm max}(H)=\frac{4\pi}{H^2},
\end{equation}
and the surface-tension term uses the excess area of the cap relative to its projected flat disk,
$\Delta A=A^2H^2/4\pi$. This landscape is an effective description rather than the microscopic KMC
Hamiltonian: lattice degrees of freedom enter only through the instantaneous $\kappa_{\rm cl}^{\rm eff}$
and $H_{\rm min}^{\rm eff}$. Normalizing energies by the bending energy $4\pi\kappa_{\rm M}$ of a
closed membrane sphere, the dimensionless landscape reads
\begin{equation}
\begin{split}
    \mathcal{E}(H,A)=&
    \frac{AH^2}{4\pi}
    +\frac{\sigma}{\kappa_{\rm M}}\frac{A^2H^2}{(4\pi)^2}\\
    +&\frac{\kappa_{\rm cl}^{\rm eff}}{4\pi\kappa_{\rm M}}
      (H-H_{\rm min}^{\rm eff})^2A-\frac{\Delta E_{\rm pol}}{4\pi\kappa_{\rm M}}\frac{A}{A_{\rm cl}},
\end{split}
\end{equation}
with the four terms representing membrane bending, membrane tension, clathrin bending, and clathrin
polymerization. Within a single instantaneous landscape $\kappa_{\rm cl}^{\rm eff}$ and
$H_{\rm min}^{\rm eff}$ are fixed; along a trajectory they evolve as the coat stiffens and develops
curvature memory.

At fixed area, $\partial_H\mathcal{E}=0$ gives the minimizing curvature
\begin{equation}
    H^*(A)=
    \frac{4\pi\kappa_{\rm cl}^{\rm eff}H_{\rm min}^{\rm eff}}
    {A\sigma+4\pi(\kappa_{\rm cl}^{\rm eff}+\kappa_{\rm M})},
\end{equation}
and substitution yields the effective one-dimensional energy
$\mathcal{E}^*(A)=\mathcal{E}(H^*(A),A)$. A finite-area basin exists only when the polymerization
gain lies within the window (cf. Fig.~\ref{fig:EnergyLandscape})
\begin{equation}
\begin{split}
    \Delta E_{\rm min}&<\Delta E_{\rm pol}<\Delta E_{\rm max},\\
    \Delta E_{\rm max}&=A_{\rm cl}(H_{\rm min}^{\rm eff})^2\kappa_{\rm cl}^{\rm eff},\\
    \Delta E_{\rm min}&=
    \frac{A_{\rm cl}(H_{\rm min}^{\rm eff})^2
    \kappa_{\rm cl}^{\rm eff}}
    {1+\kappa_{\rm cl}^{\rm eff}/\kappa_{\rm M}},
\end{split}
\end{equation}
with corresponding critical curvature and area
\begin{equation}
    H_{\rm crit}
    =
    \frac{
    \sqrt{A_{\rm cl}(H_{\rm min}^{\rm eff})^2
    \kappa_{\rm cl}^{\rm eff}-\Delta E_{\rm pol}}}
    {\sqrt{A_{\rm cl}}\sqrt{\kappa_{\rm M}+\kappa_{\rm cl}^{\rm eff}}},
\end{equation}
\begin{equation}
    A_{\rm crit}
    =
    4\pi\ell_\sigma^2
    \left[
    \frac{\kappa_{\rm cl}^{\rm eff}}{\kappa_{\rm M}}
    \left(\frac{H_{\rm min}^{\rm eff}}{H_{\rm crit}}-1\right)-1
    \right],
\end{equation}
where $\ell_\sigma=\sqrt{\kappa_{\rm M}/\sigma}$. These serve as coarse-grained indicators of
whether the moving minimum lies inside the accessible state space or beyond the closure boundary,
not as sharp microscopic fate boundaries for individual stochastic KMC trajectories. The full
derivation is given in Supplementary Note 1.5.

\subsubsection*{Deriving geometric stiffening and curvature memory}
We model the coat as a continuous small circular patch on a sphere of radius $R$, i.e. of curvature
$H=1/R$, corresponding to growth at constant curvature. A curvature strain $e=-\Delta H/H$, with
$\Delta H=H'-H$, deforms the patch; the resulting displacement field $u_\alpha$ relates to the
Cauchy strain tensor $u_{\alpha\beta}$ through
\begin{equation}
    u_{\alpha\beta}=\frac{1}{2}\left(\nabla_\alpha u_\beta+\nabla_\beta u_\alpha\right)+e g_{\alpha\beta}\,.
\end{equation}
The last term is the essential ingredient: a change in radius uniformly rescales all distances even
when no point moves relative to its surface coordinates. This term can be relaxed away in a flat
geometry, but not on the sphere. Its trace $\Xi:=\mathrm{tr}(u_{\alpha\beta})$ measures the local
area change, i.e. compression or dilation.

With the in-plane elastic energy density
\begin{equation}
    \mathcal{U}=\frac{K}{2}\Xi^2+\mu \Tilde{u}_{\alpha\beta}\Tilde{u}^{\alpha\beta}\,,
\end{equation}
where $\Tilde{u}_{\alpha\beta}:=u_{\alpha\beta}-\frac{\Xi}{2}g_{\alpha\beta}$ is the trace-free
strain and $K$, $\mu$ the two elastic moduli, the equilibrium condition
$\nabla_\beta\sigma^{\alpha\beta}=0$ for the stress
$\sigma_{\alpha\beta}:=\partial\mathcal{U}/\partial u_{\alpha\beta}$ yields the elastostatic
equation
\begin{equation}
    K\nabla^\alpha\Xi+\mu\Delta u^{\alpha}+\frac{\mu}{R^2}u^{\alpha}=0\,.
\end{equation}
Its divergence gives an inhomogeneous Helmholtz equation for $\Xi$, with an inhomogeneity intrinsic
to the curved geometry,
\begin{equation}
    \left[\Delta +\frac{1}{\lambda^2}\right]\Xi=\frac{2e}{\lambda^2}
    \quad\text{with}\quad\lambda:=R\sqrt{\frac{K+\mu}{2\mu}}\,.
    \label{eq:inhomogenous helmholtz equation}
\end{equation}
In the flat limit with azimuthal symmetry, $\Xi\equiv\Xi(r)$ and
$\Delta\approx\partial_r^2+r^{-1}\partial_r$, the general solution of
Eq.~\ref{eq:inhomogenous helmholtz equation} is
\begin{equation}
    \Xi(r)=2e+C_1J_0(r/\lambda)\,,
\end{equation}
with $J_0$ the zeroth-order Bessel function of the first kind and $C_1$ fixed by the boundary
condition of vanishing radial stress at the patch edge, $\sigma_{rr}(r_{\rm edge})=0$. This gives
\begin{equation}
    C_1\approx -2e\left[1+\frac{r_{\rm edge}^2}{4\lambda^2}\right]\,.
\end{equation}
In the small-patch limit $J_0(r/\lambda)\approx 1$, so the leading-order curvature dependence of
$\Xi$ is carried by the integration constant alone,
\begin{equation}
    \Xi\approx -\frac{e r_{\rm edge}^2}{2\lambda^2}\approx -4\frac{\mu}{K}e\phi\,,
\end{equation}
using $\pi r_{\rm edge}^2\approx A$, $\phi=AH^2/4\pi$, and $\lambda^2\approx K/2\mu H^2$. The detailed calculation can be found in Supplementary note 1.2.

We then split the energy density into a dilation contribution
$\mathcal{U}_{\rm dil}:=\tfrac{K}{2}\Xi^2$ and a bending contribution of Helfrich
form~\cite{helfrichElasticPropertiesLipid1973}, $\mathcal{U}_{\rm bend}\approx\mu(H'-H_{\rm min}^{\rm micro})^2$, where
$H'$ is the new curvature and $H_{\rm min}^{\rm micro}$ is the microscopically preferred curvature. Rearranging,
\begin{equation}
    \begin{split}
        \frac{\mathcal{U}}{L_0^2}=&\mu (H'-H_{\rm min}^{\rm micro})^2+\frac{8\mu^2}{KL_0^2}\frac{\phi^2}{H^2}(H'-H)^2\\
        =&(1+w)\mu\left[H'-\frac{H_{\rm min}^{\rm micro}+wH}{1+w}\right]^2\\
        &\quad\text{with}\quad w:=\frac{8\mu}{KL_0^2}\frac{\phi^2}{H^2}\,,
    \end{split}
\end{equation}
which recovers a standard Helfrich bending energy with the effective parameters
\begin{equation}
    \begin{split}
        \kappa^{\rm eff}:=&(1+w)\mu \approx \exp(w)\mu\,,\\
        H_{\rm min}^{\rm eff}:=&\frac{H_{\rm min}^{\rm micro}+wH}{1+w}\,.
    \end{split}
\end{equation}
For details of the derivation see Supplementary Note 1.2 and 1.3.

\clearpage 

\bibliography{literature}
\bibliographystyle{abbrv}

\section*{Acknowledgments}
We thank G. Tagiltsev and S. Scheuring for providing the rendering for Fig. 1A and the experimental data presented in Fig. 5E. We thank F. Ziebert, F. Frey and N. Winkler for insightful discussions. We acknowledge support by the Deutsche Forschungsgemeinschaft (DFG, German Research Foundation) through Priority Programme 2332 (Projektnummer 492010213)
and SFB 1638 (Projektnummer 511488495). 
We acknowledge support by the state of Baden-Württemberg through bwHPC and the DFG through grant INST 35/1597-1 FUGG, and the scientific data storage service SDS@hd supported by the Ministry of Science, Research and the Arts Baden-Württemberg (MWK) and the DFG through grant INST 35/1503-1 FUGG.

\noindent AI-assisted technologies: ChatGPT [OpenAI] and Claude [Anthropic] were used solely to improve English wording and readability of author-written text. Coding assistants [GitHub Copilot, Codex \& Claude Code] were used to assist with code writing, debugging, and/or refactoring. All AI-assisted text and code were reviewed, edited, tested, and validated by the authors, who take full responsibility for the content, analyses, and results.

\paragraph*{Author contributions:}
U.S.S. and L.L. conceived and supervised the study. J.H.H.D. and L.L. developed the numerical simulation, analysis, and visualization code, and ran the simulations. J.H.H.D. and L.L. conceived the theoretical framework, J.H.H.D. performed the mathematical derivations. J.H.H.D., U.S.S. and L.L. wrote the paper. All authors reviewed and approved the final version of the paper.

\paragraph*{Competing interests:}
There are no competing interests to declare.

\paragraph*{Data and materials availability:} The complete simulation code including documentation and the scripts and parameter table to rerun the analysis shown in this paper can be found in the following GitHub repository, after the paper is accepted for publication \url{https://github.com/Dreckhoff/ClathCoatEvolutionCode}. A detailed description of the repository is also provided in Supplementary Note 2.
The original data files for the simulation results are available from the authors upon reasonable request.

\end{document}


\onecolumn
\normalsize
\linespread{1.5}\selectfont
\newgeometry{margin=1in}

\date{}

\renewcommand{\thefigure}{S\arabic{figure}}
\renewcommand{\thetable}{S\arabic{table}}
\renewcommand{\theequation}{S\arabic{equation}}
\renewcommand{\thepage}{S\arabic{page}}

\setcounter{figure}{0}
\setcounter{table}{0}
\setcounter{equation}{0}
\setcounter{page}{1}

\begin{center}
  {\large\bfseries Supplementary Materials for}\\[0.5em]
  {\large\bfseries \scititle}\\[1.5em]
  J.\,H.\,H. Dreckhoff,
  U.\,S. Schwarz$^{\ast}$,
  L. Lettermann$^{\ast}$\\[0.4em]
  \small BioQuant, Heidelberg University, 69120 Heidelberg, Germany.\\
  \small Institute for Theoretical Physics, Heidelberg University, 69120 Heidelberg, Germany.\\[0.4em]
  \small$^\ast$Corresponding authors.\\
  \small Emails: schwarz@thphys.uni-heidelberg.de and lettermann@uni-heidelberg.de
\end{center}

\bigskip\noindent

\noindent This PDF file includes:\\
Supplementary Note \ref{sec:Calculations}: Calculations\\
Supplementary Note \ref{sec:Simulations}: Simulation code and reproducibility workflow\\
Captions for Movies S1 to S6\\

\noindent Other Supplementary Materials for this manuscript include:\\
Movies S1 to S6\\
Code repository will be available after acceptance of the manuscript.

\section{Calculations}\label{sec:Calculations}

\subsection{Calculating the microscopic clathrin bending rigidity and spring constants\label{methods:kappa calc}}

This derivation follows closely \cite{jinRigidityTriskelionArms2000}, for general information
refer to the classic \cite{lifsicTheoryElasticity1986} and \cite{gittesFlexuralRigidityMicrotubules1993}.

For the transition of bending rigidities we need a value for the individual, single-clathrin bending
rigidity. However, in experiment, usually complete clathrin cages are measured. For single clathrin, one
observes individual clathrin in solution and tracks its movements.

To model the movement of an individual clathrin arm, we use the theory of elastic rods. There, the
bending energy $E_{\rm bend}$ of a rod is given by
\begin{equation}
    E_{\rm bend}=\frac{1}{2}EI\int_0^{s_{\rm tot}}\left[C(s)-C_0(s)\right]^2\text{d}s,
\end{equation}where $C(s)$ is the curvature at arclength $s$, $C_0(s)$ is the preferred 
curvature, and $EI$ is called the flexural rigidity with units of length times energy.

Measuring the shape function $\Xi$ and decomposing it into normal modes, $\Xi(s)
=\frac{1}{\sqrt{2L}}\sum_{n=0}^\infty a_n\cos(n\pi s/L)$, one can extract the 
flexural rigidity through measurements of the mean square modes,
\begin{equation}
  \langle (a_n-a_n^0)^2\rangle=\frac{k_{\rm B}T}{EI}\left(\frac{L}{n\pi}\right)^2\,,
\end{equation}which yields $EI_{\rm arm}\approx 35\,\kbt\,\text{nm}$ for a single clathrin 
arm.

In the clathrin coat however, multiple arms bind together to form a bond, at least two, and at 
most four. For an isotropic and elastic substance, the flexural rigidity may be 
decomposed as $EI=E\cdot I$, where $E$ is the Young's modulus and $I$ is the moment
of inertia. Therefore, the combined flexural rigidity of the clathrin arms is\begin{equation}
  EI_{\rm bond}=E_{\rm clathrin}I_{\rm bond}=E_{\rm clathrin}\left(g I_{\rm arm}\right)=
  gEI_{\rm arm}
\end{equation}where $g$ is a factor depending on the geometric arrangement and number of the 
arms. Jin and Nossal present different configurations of arms with their resulting factor 
of $g$, with $g\in [12,26]$ with an average of $g\approx 16$, i.e., four bonds do not make the
arm four times as stiff, but $4^2=16$ times \cite{jinRigidityTriskelionArms2000}.

We can now relate the microscopic flexural rigidity to the bending rigidity of a patch
of clathrin to find the single-clathrin analogue to the full-cage bending rigidity. For 
this, assume we have a patch of clathrin which we want to bend. Then the usual bending energy is\begin{equation}
  E_{\rm bend, membrane}=\frac{A \kappa_{\rm cl}}{R^2}
\end{equation}In the other picture, talking about bent rods, two rods need to be bent over 
their whole length $L$ at curvature $C=1/R$ to achieve the same overall curvature. There, the
bending energy of a single grid cell is\begin{equation}
  E_{\rm bend, rod}=2\frac{1}{2}EI_{\rm bond}\int_0^LC(s)^2\text{d}s=\frac{EI_{\rm bond}L}{R^2}\,.
\end{equation}
For a single grid cell, with $A_{\rm grid cell}\approx L^2$, both equations should match, resulting
in\begin{equation}
  \frac{L^2 \kappa_{\rm cl}}{R^2}=\frac{EI_{\rm bond}L}{R^2}\quad\Rightarrow\quad\kappa_{\rm cl}=
  g\frac{EI_{\rm arm}}{L}
\end{equation}For a grid cell, $L\approx 18.4\,\text{nm}$, resulting in $EI_{\rm arm}/L\approx 1.9\,\kbt$.
Depending on the assumptions taken, this leaves room for $\kappa_{\rm cl}<52\,\kbt$. Averaging over the involved legs we end up with more realistic values around $\kappa_{\rm cl}\in [6,13]\,\kbt$.

It is of course difficult to assign a precise value. The important note however is that
this value is far below the reported $100-400\,\kbt$ for the full cage bending rigidity!

Related to the single-clathrin bending rigidity is the calculation of the constants for the 
harmonic potentials governing our microscopic dynamics. For the angles $\varphi$ and $\chi$, 
as well as for the length of the legs $L$, we use harmonic potentials with spring constants
$k_X$ (with $X=\varphi, \chi, L$) to constrain the movement.

For convenience, assume that $X$ is given dimensionless and in reference to a preferred
value $X_0$, i.e. the potential has the form\begin{equation}
  E_X(\Delta X)=k_X(\Delta X)^2\,,
\end{equation}where we keep with the convention of no factor of $1/2$ in the spring constant
for consistency, and $\Delta X=X-X_0$. Let $\Delta X_{\rm max}$ be the deviation that
is realistically expected to occur. The energy of such a configuration (including the polymerization
energy per leg, $E_{\rm pol. leg}\approx 5\,\kbt$) has to be on the 
order of the temperature to not be suppressed:\begin{equation}
  E(\Delta X_{\rm max})-E_{\rm pol. leg} \approx \kbt=\frac{1}{\beta}\quad\Rightarrow
  \quad k_X\approx \frac{1+\beta E_{\rm pol. leg}}{\beta \Delta X_{\rm max}^2}\,.
\end{equation}In units of $\kbt$, $\beta=1$. Again from Jin and Nossal \cite{jinRigidityTriskelionArms2000}, the
typical order of expected deviations can be taken. For a maximal deviation of 
$\Delta \varphi_{\rm max}\approx 0.6\,\text{rad}$ we expect $k_{\varphi}$ on the order
of $\approx 20\,\kbt$. For
a maximal deviation of $\Delta L_{\rm max}\approx 0.1\,\text{L}_0$ we expect $k_L$ on the
order of $\approx 600\,\kbt$.

We identify $k_{\varphi}$ with the spring constant for angular deviations, and $k_L$ with the spring
constant for length deviations. From our experience, we achieved more regular
results with a bit higher of a stretching constant, which is why we used a slightly
larger value of $k_L=800\,\kbt$.

\subsection{Calculating the dilation field and dilation energy\label{methods:dilation field}}
We want to derive how a lattice in a spherical geometry reacts to changes 
in said geometry. The central result is an analytic expression for the
 dilation field (measuring how much the lattice is locally compressed or stretched) that 
 accumulates when
a coat assembled at curvature $H$ -- the growth curvature -- is transformed to 
a curvature $H'$, and how
this dilation field gives rise to the geometry-induced 
coat stiffening reported in the
main text.

\subsubsection*{Setup and assumptions.}
We treat the clathrin coat as a 2D elastic continuum on a spherical surface of radius $R$,
using surface coordinates $(r,\varphi)$ where $r$ is the geodesic distance from the pole.
The metric is diagonal: $g_{rr}=1$ and $g_{\varphi\varphi}=R^2\sin^2(r/R)$, and we will be working in standard Riemannian geometry.
In the following we list our assumption for the succeeding derivation:

\begin{enumerate}
  \item[(A1)] \textit{2D continuum elasticity.}
    The coat is described by a shear modulus $\mu$ with $[\mu]=\kbt/\text{L}_0^2$ and a bulk modulus $K$ with $[K]=\kbt/\text{L}_0^2$. Both are related to the microscopic spring constants $k_\text{bend}$ and $k_\text{stretch}$ describing the microscopic clathrin deformations, with $\mu=k_\text{bend}/\text{L}_0^2$ and $K=k_\text{stretch}/\text{L}_0^2$. The shear modulus is a measure of resistance of the coat to shear forces, which is caused by the microscopic angular bending stiffness of the clathrin legs. As a reminder, the typical quantities are $k_\text{bend}\sim 20\,\kbt$ and $k_\text{stretch}\sim 800\,\kbt$, thereby $k_\text{bend}/k_\text{stretch}=\mu/K\ll 1$.
  \item[(A2)] \textit{Small strain.}We are working in the limit of small curvature changes, i.e., small strain $|e| = |\Delta H/H| \ll 1$; linearised (Cauchy) elasticity applies.
  \item[(A3)] \textit{Small patch.}We will look at a clathrin patch that, at curvature $H$, only covers a small part of the potential sphere, i.e., the closure $\phi=AH^2/4\pi\ll 1$. Therefore the patch radius satisfies $r_{\rm edge} \ll R \ll \lambda$, where
    $\lambda = R\sqrt{(K+\mu)/2\mu}\approx R\sqrt{K/2\mu}$ is the elastic correlation
    length.
  \item[(A4)] \textit{Stress-free boundary.}
    The radial stress $\sigma_{rr}$ vanishes at $r=r_{\rm edge}$, corresponding to a
    freshly incorporated ring that is mechanically relaxed at its edge.
\end{enumerate}

\subsubsection*{Strain tensor and elastostatic equilibrium.}Imagine a situation where 
a spherical patch exists at curvature $H$, and now the curvature is changed to $H'$. 
A change from curvature $H$ to $H'=H+\Delta H$ induces a homogeneous strain
$e:=\Delta R/R = -\Delta H/H + \mathcal{O}((\Delta H/H)^2)$ on the lattice. 
Let $u_\alpha$ be the displacement field on the sphere, measuring the displacement of 
any point before and after the curvature change.
The Cauchy strain tensor, describing the \textit{relative displacement between 
neighboring points} (i.e., how much points are locally shifted) is given by
\begin{equation}
  u_{\alpha\beta} \;:=\;
    \tfrac{1}{2}(\nabla_\alpha u_\beta + \nabla_\beta u_\alpha) + e\,g_{\alpha\beta}\,.
  \label{eq:cauchy}
\end{equation}The last summand is what changes compared to a flat geometry. Because different 
spherical geometries are incompatible,~\cite{efratiElasticTheoryUnconstrained2009} a uniform displacement
is forced upon all points. However, a priori this does not need to induce stresses, since the grid 
could still relax to mitigate this dilation (which is what happens in the flat geometry). 

The Cauchy tensor 
can be decomposed into its trace $\Xi:=g^{\alpha\beta}u_{\alpha\beta}=
\nabla_\alpha u^\alpha + 2e$ (measuring volume change) and the
trace-free shear part $\tilde{u}_{\alpha\beta}$ (measuring shape change),
\begin{equation}
  u_{\alpha\beta} = \tfrac{\Xi}{2}\,g_{\alpha\beta} + \tilde{u}_{\alpha\beta}\,.
  \label{eq:decomp}
\end{equation}
Then, the in-plane elastic energy density is given by
\begin{equation}
  \mathcal{U} = \tfrac{K}{2}\,\Xi^2 + \mu\,\tilde{u}_{\alpha\beta}\tilde{u}^{\alpha\beta}
             = \mu\,u_{\alpha\beta}u^{\alpha\beta} + \tfrac{K-\mu}{2}\,\Xi^2\,,
  \label{eq:energy_density}
\end{equation}
which is the standard isotropic elastic energy for a 2D material.
The stress tensor $\sigma^{\alpha\beta} := \partial\mathcal{U}/\partial u_{\alpha\beta}
= 2\mu u^{\alpha\beta}+(K-\mu)g^{\alpha\beta}\Xi$ in equilibrium must satisfy $\nabla_\beta\sigma^{\alpha\beta}=0$, describing force balance at every point (no stress divergence), giving the
elastostatic equation
\begin{equation}
  K\nabla^\alpha\Xi \;+\; \mu\,\Delta u^\alpha \;+\; \frac{\mu}{R^2}\,u^\alpha \;=\; 0,
  \label{eq:elastostatic}
\end{equation}
since\begin{equation}
    \begin{split}
        \nabla_\beta \left(g^{\alpha\beta}\Xi\right)=g^{\alpha\beta}\nabla_\beta\Xi
    \end{split}
\end{equation}in Riemannian geometry as $\nabla_\beta g^{\alpha\beta}=0$ and \begin{equation}
    \begin{split}
        \nabla_\beta u^{\alpha\beta}=&\frac{1}{2}\nabla_\beta\left[\nabla^\alpha u^\beta+\nabla^\beta u^\alpha + \frac{\Delta R}{R}g^{\alpha\beta}\right]\\
        =&\frac{1}{2}\left[\nabla_\beta\nabla^\alpha u^\beta+\Delta u^\alpha\right]\\
        =&\frac{1}{2}\left(\left[\nabla^\alpha\nabla_\beta +R^{\alpha}_{\,\beta}\right]u^\beta + \Delta u^\alpha\right)\\
        =&\frac{1}{2}\left(\nabla^\alpha \Xi + \frac{1}{R^2}u^\alpha + \Delta u^\alpha\right)
    \end{split}
\end{equation}where we used the Ricci-identity for swapping the covariant derivatives, used that the Ricci tensor in our coordinates is $R^\alpha_{\,\beta}=\delta^\alpha_\beta/R^2$. and used  that $\nabla^\alpha (\nabla_\beta u^\beta)=\nabla^\alpha(\nabla_\beta u^\beta + 2\Delta R/R)=\nabla^\alpha \Xi$.

\subsubsection*{Explicit divergence of the elastostatic equation.}
We now spell out the step in which a second covariant divergence is applied to
Eq.~(\ref{eq:elastostatic}).  Acting with $\nabla_\alpha$ gives
\begin{equation}
  K\Delta\Xi
  + \mu\,\nabla_\alpha\Delta u^\alpha
  + \frac{\mu}{R^2}\nabla_\alpha u^\alpha
  =0.
  \label{eq:div_elastostatic_start}
\end{equation}
The only non-trivial term is the divergence of the vector Laplacian.  On a
sphere of constant radius $R$, commuting the derivatives once gives
\begin{equation}
  \nabla_\alpha\Delta u^\alpha
  = \nabla_\alpha\nabla_\beta\nabla^\beta u^\alpha
  = \Delta(\nabla_\alpha u^\alpha)
    + R_{\alpha\beta}\nabla^\alpha u^\beta .
  \label{eq:div_vector_laplacian_identity}
\end{equation}
Here $R_{\alpha\beta}=g_{\alpha\beta}/R^2$, so the curvature term is simply
\begin{equation}
  R_{\alpha\beta}\nabla^\alpha u^\beta
  = \frac{1}{R^2}g_{\alpha\beta}\nabla^\alpha u^\beta
  = \frac{1}{R^2}\nabla_\alpha u^\alpha .
  \label{eq:ricci_divergence_term}
\end{equation}
With $q:=\nabla_\alpha u^\alpha$, Eq.~(\ref{eq:div_elastostatic_start})
therefore becomes
\begin{equation}
  K\Delta\Xi
  + \mu\left(\Delta q + \frac{q}{R^2}\right)
  + \frac{\mu}{R^2}q
  =0.
  \label{eq:div_elastostatic_q}
\end{equation}
The trace definition $\Xi=\nabla_\alpha u^\alpha+2e$ implies
$q=\Xi-2e$.  Since the imposed geometric strain $e=\Delta R/R$ is spatially
constant for a homogeneous change of sphere radius, $\Delta q=\Delta\Xi$.
Substituting this into Eq.~(\ref{eq:div_elastostatic_q}) gives
\begin{equation}
  (K+\mu)\Delta\Xi
  + \frac{2\mu}{R^2}(\Xi-2e)=0.
  \label{eq:div_elastostatic_theta}
\end{equation}
Dividing by $K+\mu$ and defining
$\lambda^{-2}:=2\mu/[R^2(K+\mu)]$ yields the
\emph{inhomogeneous Helmholtz equation} for the dilation field:
\begin{equation}
  \left[\Delta + \frac{1}{\lambda^2}\right]\Xi \;=\; \frac{2e}{\lambda^2},
  \qquad
  \lambda \;:=\; R\sqrt{\frac{K+\mu}{2\mu}} \;\approx\; R\sqrt{\frac{K}{2\mu}}\,.
  \label{eq:helmholtz}
\end{equation}

\subsubsection*{Solution in the small-patch limit.}
Assuming azimuthal symmetry ($\Xi\equiv\Xi(r)$) and invoking assumption (A3), the small patch size, the
Laplacian can be approximated by its flat-space form $\Delta\approx\partial_r^2+r^{-1}\partial_r$, since the essential feature of spherical geometry is the inhomogeneity.
The general solution of Eq.~(\ref{eq:helmholtz}) is
\begin{equation}
  \Xi(r) = 2e + C_1 J_0(r/\lambda),
  \label{eq:theta_general}
\end{equation}
where $J_0$ is the zeroth-order Bessel function of the first kind and $C_1$ is set by
condition (A4).

Using the azimuthal symmetry, the only non-vanishing Christoffel symbols (reminder: $\nabla_\alpha u^\beta:=\partial_\alpha u^\beta+\Gamma_{\alpha\lambda}^\beta u^\lambda$ and $\nabla_\alpha u_\beta:=\partial_\alpha u_\beta - \Gamma_{\alpha\beta}^\lambda u_\lambda$) are $\Gamma_{\varphi\varphi}^r = -R\sin(r/R)\cos(r/R)$ and
$\Gamma_{r\varphi}^{\varphi}=\cos(r/R)/R$. Therefore we find\begin{equation}
    u_{rr}=\nabla_ru_r+eg_{rr}=\partial_ru_r+e
\end{equation}and\begin{equation}
    u_{\phi\phi}=\nabla_\phi u_\phi +eg_{\phi\phi}=R\sin\left(\frac{r}{R}\right)\cos\left(\frac{r}{R}\right)u_r(r)+eR^2\sin^2\left(\frac{r}{R}\right)
\end{equation}The dilation, as the trace of the Cauchy tensor, is simply the sum of both terms combined using the inverse metric, leading to\begin{equation}
    \begin{split}
        \Xi=\text{tr}(u_{\alpha\beta})=&g^{\alpha\beta}u_{\alpha\beta}\\
        =&\partial_ru_r+e+\frac{1}{R^2\sin^2\left(\frac{r}{R}\right)}u_{\phi\phi}\\
        =&2e+\partial_ru_r+\frac{u_r(r)}{R}\cot\left(\frac{r}{R}\right)\\
        \approx&2e+\partial_ru_r+\frac{u_r(r)}{r} \;\;@\;\;r\ll R
    \end{split}
\end{equation}Using $\partial_r(ru_r)/r=\partial_ru_r+u_r/r$, we find\begin{equation}
    \Xi=2e+\frac{1}{r}\partial_r(u_rr)\,.
\end{equation}Using the solution for $\Xi(r)=2e+C_1J_0(r/\lambda)$, we find\begin{equation}
    \frac{1}{r}\frac{\text{d}}{\text{d}r}\left(ru_r\right)=C_1J_0(r/\lambda)\Rightarrow \frac{r u_r}{C_1}=\int \text{d}r\,rJ_0(r/\lambda)=\lambda r J_1(r/\lambda)+C_2
\end{equation}using
$\int x J_0(x)\,\mathrm{d}x=xJ_1(x)$. Finiteness at $r=0$ dictates $C_2=0$, giving us\begin{equation}
    u_r(r)=\lambda C_1 J_1(r/\lambda)
\end{equation}
The radial stress at the patch edge is \begin{equation}
    \begin{split}
        \sigma_{rr}=&2\mu u_{rr}+(K-\mu)\Xi\\
        =&2\mu \left(\partial_ru_r+e\right)+(K-\mu)\left(2e+C_1J_0(r/\lambda)\right)\\
        =&2\mu\left(\partial_r(\lambda C_1J_1(r/\lambda))+e\right)+(K-\mu)(2e+C_1J_0(r/\lambda))\\
        =&2\mu \left[C_1\left(J_0(r/\lambda)-\frac{\lambda}{r}J_1(r/\lambda)\right)+e\right]+(K-\mu)(2e+C_1J_0(r/\lambda))\\
        =&2Ke+C_1\left[(K+\mu)J_0(r/\lambda)-\frac{2\mu\lambda}{r}J_1(r/\lambda)\right]
    \end{split}
\end{equation}Then
\begin{equation}
  \sigma_{rr}(r_{\rm edge}) = 2Ke + C_1\!\left[
    (K+\mu)\,J_0^{\rm edge} - \frac{2\mu\lambda}{r_{\rm edge}}\,J_1^{\rm edge}
  \right],
  \label{eq:sigma_rr}
\end{equation}
where $J_n^{\rm edge}:=J_n(r_{\rm edge}/\lambda)$.
Setting $\sigma_{rr}(r_{\rm edge})=0$ and expanding in $r_{\rm edge}/\lambda\ll 1$
(assumption A3) gives $C_1\approx-2e[1+r_{\rm edge}^2/4\lambda^2]$. For a qualitative understanding of the behaviour, it suffices to substitute back into Eq.~(\ref{eq:theta_general}) and retain the leading order in
$r_{\rm edge}/\lambda$,
\begin{equation}
  \Xi \;\approx\; -\frac{e\,r_{\rm edge}^2}{2\lambda^2}
  \qquad (r \ll R \ll \lambda).
  \label{eq:theta_intermediate}
\end{equation}
Using $\pi r_{\rm edge}^2\approx A$ and $\phi=AH^2/4\pi$, the geometric prefactor is
rewritten in terms of closure to give the key result:
\begin{equation}
  \boxed{
    \Xi(\phi) \;\approx\; -4\,\frac{\mu}{K}\,e\,\phi.
  }
  \label{eq:theta_phi}
\end{equation}

\noindent Equation~(\ref{eq:theta_phi}) has a transparent physical interpretation:
a curvature change produces a total strain $e$; most of it is absorbed by cheap in-plane
shear (modulus $\mu$), but a residual dilation proportional to $\phi$ cannot be
accommodated because the compact spherical geometry progressively constrains planar
rearrangements as coverage grows.
For $\phi\to 0$ (flat plaque), $\Xi\to 0$: shear suffices and no stiffening occurs.

\subsubsection*{Dilation energy and stiffening.\label{sec:dilation energy and stiffening}} For a patch undilated at closure $\phi$, the energy density from the bulk modulus $\mathcal{U}_{\rm dil}=\frac{K}{2}\Xi^2$ can be used to construct an effective energy. We first consider the case where the connected coat is grown at a single curvature $H$. This is the fixed-growth-curvature protocol used in Fig.~2 of the main text.

We combine the dilation energy with the shear deformation energy from $\mu$, expressed in the 
usual Helfrich-type form using $\kappa=\mu L_0^2$,
\begin{equation}
    \Xi(H';H)=-4\frac{\mu}{K}\,e\,\phi,
    \qquad
    e=-\frac{H'-H}{H}.
    \label{eq:theta_eff_locked}
\end{equation}
Using this connected-coat amplitude for the stiffness matching gives the dilation contribution
\begin{equation}
\begin{split}
    \frac{\mathcal{U}_{\rm dil}}{L_0^2}
    &=\frac{K}{2L_0^2}\Xi^2(H';H)\\
    &=\frac{8\mu^2}{K L_0^2}\frac{\phi^2}{H^2}(H'-H)^2.
\end{split}
\label{eq:U_dil}
\end{equation}
The total effective energy is therefore
\begin{equation}
    \frac{\mathcal{U}_{\rm tot}}{L_0^2}
    =\mu(H'-H_{\rm min}^{\rm micro})^2+\frac{\mathcal{U}_{\rm dil}}{L_0^2}.
\end{equation}
Substituting Eq.~(\ref{eq:U_dil}) yields
\begin{equation}
\begin{split}
    \frac{\mathcal{U}_{\rm tot}}{L_0^2}
    =&\mu(H'-H_{\rm min}^{\rm micro})^2
      +\frac{8\mu^2}{K L_0^2}\frac{\phi^2}{H^2}(H'-H)^2\\
    =&\mu\left[(H')^2-2H'H_{\rm min}^{\rm micro}+{H_{\rm min}^{\rm micro}}^2
      +\frac{8\mu}{K L_0^2}\frac{\phi^2}{H^2}
      \left((H')^2-2H'H+H^2\right)\right]\\
    =&\mu\left[(1+w)(H')^2-2H'(H_{\rm min}^{\rm micro}+wH)+{H_{\rm min}^{\rm micro}}^2+wH^2\right]\\
    =&(1+w)\mu\left[H'-\frac{H_{\rm min}^{\rm micro}+wH}{1+w}\right]^2+\mathrm{const.}
\end{split}
\end{equation}
where
\begin{equation}
    \boxed{w:=\frac{8\mu}{K L_0^2}\frac{\phi^2}{H^2}
    =\frac{2\mu}{\pi K L_0^2}A\phi
    =\frac{\mu}{2\pi^2 K L_0^2}A^2H^2:=C\,A^2H^2.}
    \label{eq:w_fixed_corrected}
\end{equation}
Thus
\begin{equation}
    \boxed{C=\frac{\mu}{2\pi^2K L_0^2}.}
    \label{eq:C_corrected}
\end{equation}
The effective bending rigidity and preferred curvature are therefore
\begin{equation}
    \boxed{\kappa^{\rm eff}=(1+w)\kappa\approx\kappa\,\exp(w)}
    \label{eq:kappa_eff_fixed_corrected}
\end{equation}
and
\begin{equation}
    \boxed{H_{\rm min}^{\rm eff}=\frac{H_{\rm min}^{\rm micro}+wH}{1+w}=H_{\rm min}^{\rm micro}+\frac{w}{1+w}(H-H_{\rm min}^{\rm micro}).}
    \label{eq:H0_eff_fixed_corrected}
\end{equation}
By defining \begin{equation}
  \omega:=\frac{w}{1+w}
\end{equation}we see that the effective curvature is
 a weighted average of the original preferred curvature and
  the current curvature, \begin{equation}
    \boxed{H_{\rm min}^{\rm eff}=(1-\omega)H_{\rm min}^{\rm micro}+\omega H\,,}
  \end{equation}
  with the weight $\omega$
   increasing with the combined factor $A\phi$. The effective
    curvature shifts from $H_{\rm min}^{\rm micro}$ to $H$ as the coat grows and closes,
     and the stiffening grows exponentially with this same combined factor. Both effects
      are caused by the same underlying geometric constraint, which is
       captured by the single variable $w$.
For the effective energy minimizing curvature, plotting $(H_{\rm min}^{\rm eff}-H_{\rm min}^{\rm micro})/(H-H_{\rm min}^{\rm micro})$ against $w/(1+w)$ should result in a straight line from the origin with slope one. Looking at the factor $w$, we see that the stiffness does not scale with closure alone, but with the combined factor $A\phi$. The stiffening and the shift in preferred curvature are caused by both the growth of the coat and the increase of closure. A coat of any size that remains flat ($\phi=0$) does not show geometric stiffening, and a highly closed coat with very small area stiffens only weakly.

\subsection{Phenomenological extension to variable curvature during growth.}
We now extend the fixed-growth-curvature result to the case where the membrane curvature changes while new lattice material is incorporated. We do not attempt to derive the full history-dependent elastic problem here. Instead, we use the fixed-curvature result as a local building block and postulate that the connected coat can be decomposed into material increments, or ``rings'', that each contribute a quadratic dilation penalty, 
which also gives a compact notation for cutting experiments.

Let ring $i$ denote a small material increment of area $\Delta A_i$. Let $A$ be the
final coat area at which the effective stiffness is measured. The ring is
assumed to have been incorporated, or mechanically locked, when the coat had
curvature $H_i$. If the coat is later evaluated at curvature $H'$, the relative
curvature strain associated with this ring is
$e_i(H')=-(H'-H_i)/H_i$, in the same small-strain sense used above. We then
postulate the ring contribution
\begin{equation}
    \frac{\mathcal{U}_{{\rm dil}, i}}{L_0^2}
    =\frac{\Delta A_i}{A}
    \frac{8\mu^2}{K L_0^2}\phi^2
    \frac{(H'-H_i)^2}{H_i^2}
    =\mu\,w_i\,(H'-H_i)^2,
    \label{eq:ring_dilation_energy_current_phi}
\end{equation}
with the ring weight
\begin{equation}
    w_i:=
    \frac{\Delta A_i}{A}
    \frac{8\mu}{K L_0^2}
    \frac{\phi^2}{H_i^2}.
    \label{eq:w_i_variable_current_phi}
\end{equation}
Here $H_i$ is the curvature at which the material increment was incorporated,
whereas $\phi$ is the closure of the coat at the state where the effective
response is evaluated. Thus the curvature memory of each ring is set by its
birth curvature $H_i$, but the geometric stiffening amplitude is controlled by
the current compactness of the whole coat. This convention recovers the
fixed-growth-curvature result when all material increments are incorporated and
evaluated at the same curvature.

The full coat energy is
\begin{equation}
    \frac{\mathcal{U}_{\rm coat}(H')}{L_0^2}
    =\mu(H'-H_{\rm min}^{\rm micro})^2+\mu\sum_i w_i(H'-H_i)^2.
    \label{eq:ring_sum}
\end{equation}
This expression is a sum of quadratic terms in the evaluation curvature $H'$ and
can therefore be completed exactly. We define the total stiffening weight and
the normalized curvature memory as
\begin{equation}
    \mathcal{W}:=\sum_i w_i,
    \qquad
    \mathcal{H}:=\frac{\sum_i w_iH_i}{\sum_i w_i}.
    \label{eq:W_H_variable_current_phi}
\end{equation}
Substituting these definitions into Eq.~(\ref{eq:ring_sum}) gives
\begin{equation}
\begin{split}
    \frac{\mathcal{U}_{\rm coat}(H')}{L_0^2}
    &=\mu\left[(1+\mathcal{W})(H')^2
      -2H'(H_{\rm min}^{\rm micro}+\mathcal{W}\mathcal{H})
      +{H_{\rm min}^{\rm micro}}^2+\sum_i w_iH_i^2\right]\\
    &=\mu(1+\mathcal{W})
      \left[H'-\frac{H_{\rm min}^{\rm micro}+\mathcal{W}\mathcal{H}}{1+\mathcal{W}}\right]^2
      +\mathrm{const.}
\end{split}
    \label{eq:ring_sum_completed_current_phi}
\end{equation}
The variable-curvature extension therefore predicts
\begin{equation}
    \boxed{\kappa_{\rm cl}^{\rm eff}
    =\kappa_{\rm cl}(1+\mathcal{W})
    \approx\kappa_{\rm cl}\exp(\mathcal{W})}
    \label{eq:mueff}
\end{equation}
and
\begin{equation}
    \boxed{H_{\rm min}^{\rm eff}
    =\frac{H_{\rm min}^{\rm micro}+\mathcal{W}\mathcal{H}}{1+\mathcal{W}}.}
    \label{eq:H0eff}
\end{equation}

Eq.~(\ref{eq:H0eff}) can be rewritten as
\begin{equation}
    H_{\rm min}^{\rm eff}
    =H_{\rm min}^{\rm micro}+\frac{\mathcal{W}}{1+\mathcal{W}}
      \left(\mathcal{H}-H_{\rm min}^{\rm micro}\right).
    \label{eq:Hpref_memory_shift_current_phi}
\end{equation}
With
\begin{equation}
    \omega:=\frac{\mathcal{W}}{1+\mathcal{W}},
    \label{eq:omega_variable_current_phi}
\end{equation}
this becomes
\begin{equation}
    \boxed{
    H_{\rm min}^{\rm eff}
    =(1-\omega)H_{\rm min}^{\rm micro}+\omega\mathcal{H}.}
    \label{eq:Hpref_weighted_average_current_phi}
\end{equation}
Thus the effective preferred curvature is a weighted average between the
microscopic preferred curvature and the growth curvatures memorised by the
material increments. The weights are not simple area weights: through
Eq.~(\ref{eq:w_i_variable_current_phi}) they contain the current closure
$\phi$, the fractional area contribution $\Delta A_i/A$, and the curvature
scale at incorporation.

The fixed-growth-curvature result is recovered as a consistency check. If all
material increments are incorporated at the same curvature,
$H_i=H_{\rm growth}$, and the response is evaluated at that same curvature,
then
\begin{equation}
    \mathcal{W}
    =\sum_i
    \frac{\Delta A_i}{A}
    \frac{8\mu}{K L_0^2}
    \frac{\phi^2}{H_{\rm growth}^2}
    =
    \frac{8\mu}{K L_0^2}
    \frac{\phi^2}{H_{\rm growth}^2}
    =w.
    \label{eq:mueff_fixed}
\end{equation}
In the same limit, $\mathcal{H}=H_{\rm growth}$ and
Eq.~(\ref{eq:Hpref_weighted_average_current_phi}) reduces to the fixed-curvature
expression in Eq.~(\ref{eq:H0_eff_fixed_corrected}).

For comparison with simulation data, Eq.~(\ref{eq:Hpref_memory_shift_current_phi})
suggests the transformation
\begin{equation}
    \frac{H_{\rm min}^{\rm eff}-H_{\rm min}^{\rm micro}}
    {\mathcal{H}-H_{\rm min}^{\rm micro}}
    =\omega
    =\frac{\mathcal{W}}{1+\mathcal{W}}.
    \label{eq:Hpref_collapse_variable_current_phi}
\end{equation}
The exact ring-weighted growth curvature $\mathcal{H}$ is a
history-dependent quantity. When only the final growth curvature is used as a
proxy, the collapse should therefore be interpreted as a phenomenological test
of the predicted trend rather than as a direct parameter-free measurement of
Eq.~(\ref{eq:Hpref_collapse_variable_current_phi}).

\subsection{In the context of non-Euclidean plates}

While our case is very specific, the general framework we are dealing with
can be understood in the context of a series of papers by Efrati, Sharon and Kupferman 
on non-euclidean plates \cite{efratiElasticTheoryUnconstrained2009,sharonMechanicsNonEuclideanPlates2010}, 
which were preceded by the work of Klein et al. on swelling 
gels~\cite{kleinShapingElasticSheets2007}. We will now partially summarize and explain their
results in the context of our work.

When talking about deformations of shapes, one usually assumes that the shape has a rest
configuration described by its positions $\boldsymbol{x}\in \Omega$. 
After deformation, the positions are mapped to the deformed state $\boldsymbol{r}(x)$. The difference between these two
states can be quantified using the metric tensors describing each shape. For two points, $\boldsymbol{x}$ 
and $\boldsymbol{x}+\boldsymbol{\text{d}x}$, their distance initially is given by\begin{equation}
  \text{d}s^2_{\rm init}=\bar{g}_{ij}\,\text{d}x^i\,\text{d}x^j
\end{equation}where $\bar{g}$ is called the rest-metric. 
After deformation, the
point $\boldsymbol{x}$ is mapped to $\boldsymbol{r}(\boldsymbol{x})$, while $\boldsymbol{x}+\text{d}\boldsymbol{x}$
 is mapped to $\boldsymbol{r}(\boldsymbol{x}+\text{d}\boldsymbol{x})\approx \boldsymbol{r}(\boldsymbol{x})
 + \partial_i\boldsymbol{r}\,\text{d}\boldsymbol{x}^i$. The distance 
 between these two points in the deformed state is\begin{equation}
  \text{d}s^2_{\rm def}=\vert \text{d}\boldsymbol{r}\vert^2=\text{d}\boldsymbol{x}^i\text{d}\boldsymbol{x}^j \partial_i\boldsymbol{r}\cdot \partial_j\boldsymbol{r}\equiv \text{d}\boldsymbol{x}^i\text{d}\boldsymbol{x}^j g_{ij}
 \end{equation}which defined the deformed metric $g$. All deformation information, except for global
 translations and rotations, is captured in the difference of the metrics, and one defines the strain
 tensor as\begin{equation}
  \epsilon_{ij}=\frac{1}{2}(g_{ij}-\bar{g}_{ij}).
\end{equation}Often, one sees the following simplification. If we look at a 3d shape, it is always possible to
define $\bar{g}_{ij}=\delta_{ij}$, so that the rest-metric is the euclidian metric. For this, the rest-coordinates are
simply chosen to be the euclidian coordinates of the 3d embedding space. Then, we reduce to
\begin{equation}
  \epsilon_{ij}=\frac{1}{2}(g_{ij}-\delta_{ij}).
\end{equation}
Crucially, this simplification works only for a \emph{flat} rest geometry (e.g.\ a 3d body, a codimension-zero chunk of Euclidean space). For a curved 2d surface like our coat it is impossible: a sphere has no global Euclidean 2d coordinates (theorema egregium), and this very obstruction is the origin of the residual strain.

Under the assumption that the energy is only a function of strain, its lowest order expansion needs to be
of the form\begin{equation}
  \mathcal{U}=\frac{1}{2}A^{ijkl}\epsilon_{ij}\epsilon_{kl}
\end{equation}where $A^{ijkl}$ is the elastic tensor. 
With regard to symmetries and isotropy, it
is possible to find the generic form of the elastic tensor, which is where the bulk and shear
moduli enter. The important note however is that the energy is a function of the strain.

Usually, the rest-metric $\bar{g}$ is computed from a rest-state. Efrati, Sharon and Kupferman now
proposed that one can relax this constraint, and simply \textit{assume} a rest metric $\bar{g}$. Importantly, while
every shape defines a metric, not every metric defines a shape! Such metrics that define
"impossible" shapes are called non-immersible. Their example is a flat disc out of a gel that
expands upon heating. If one now only heats the center of the disc, this part wants to expand, while the
outer part does not. Every tiny patch of disk \textit{has} a rest configuration. However the collection of
patches cannot assume a shape that satisfies all rest configurations at the same time. In this case, there
exists a rest-metric $\bar{g}$, that however does not match any shape.

They mention that two very common routes where such a situation can occur are swelling, like in the
example above, or assembly at different geometries, like in our clathrin case.~\cite{sharonMechanicsNonEuclideanPlates2010} Patches of clathrin assemble
at a geometry, which defines their patch-rest-metric. However, as curvature changes, newer patches are
included at a different rest-metric. This history-dependent reference geometry is precisely the elastic ``memory'' of non-Euclidean sheets, and is what we report as the coat's \emph{curvature memory}.~\cite{sharonMechanicsNonEuclideanPlates2010} 
This can be seen quite well in our case. The only parameter
that differentiates geometries is the curvature $H$, which is related to the Gaussian curvature
$K=H^2$. The famous theorema egregium states that this curvature is an intrinsic property of the
geometry. A coat assembled at $H$ and forced to $H'$ cannot do this without stretching. Said differently, there
is no shape of the spherical cap that can match $g=\bar{g}$, and residual strain will always be present, leading
to the source term $2e/\lambda^2$ in our Helmholtz equation.

Two further remarks place our model precisely within this framework. First, because our coat is constrained to a spherical-cap geometry, it cannot relieve the curvature mismatch by buckling out of that family: the out-of-plane (buckling) branch of non-Euclidean plate theory is excluded by construction, and only the in-plane stretching response -- our dilation field -- remains. Second, the Helfrich-type bending term $\kappa H^2$ used in our energy landscape is itself the non-Euclidean-plate bending content: for an isometric (stretch-free) configuration it equals the Willmore functional, which for a spherical cap reduces to $\int H^2\,\mathrm{d}A=H^2A$.~\cite{efratiElasticTheoryUnconstrained2009}

In a newer work, Meiri and Efrati show that shapes with incompatible metrics will show a
super-extensive energy scaling with system size.~\cite{meiriCumulativeGeometricFrustration2021}. Many
systems exhibit an energy scaling $E\propto M$ that is linear in the system size, which is called
"extensive". Usually, a system consists of small units, and every unit adds some constant energy. Super-extensive
scaling means that $E\propto M^\lambda$ with $\lambda>1$. This means that every unit also increases its own
energy with system size, and is to be expected if frustration accumulates in systems with metric-incompatibility.

Assume that the strain $\epsilon=\frac{1}{2}(g-\bar{g})$ is minimized, but not zero everywhere. Then
we can expand $\epsilon$ in powers of the system size:\begin{equation}
  \epsilon=\epsilon_{00} + \epsilon_{01}x+\epsilon_{10}y+\epsilon_{11}xy+\hdots
\end{equation}
Let $\epsilon_{ij}$ be the first non-zero term in this expansion, and call $\eta:=i+j$. Then
the energy will scale like\begin{equation}
  E\sim \int\text{d}^dr\,\epsilon^2\sim r^d\,r^{2\eta}\sim M^{1+\frac{2\eta}{d}}
\end{equation}with $M\sim r^d$ being the system size. In our case, we are dealing with a
2d coat, so $d=2$ and the system size $M=A$ is the area. To find $\eta$, remember that by
definition\begin{eqnarray}
  g=\bar{g}+2\epsilon
\end{eqnarray}The constant and linear coefficients $\epsilon_{00},\epsilon_{01},\epsilon_{10}$ are compatible (they derive from a displacement) and are set to zero by energy minimization. The first coefficient that cannot be removed is fixed by the metric incompatibility, i.e.\ the Gaussian-curvature mismatch; and since $K$ is computed from the \emph{second} derivative of the metric (theorema egregium), this obstruction enters at second order. Therefore $\eta=2$ and we
find\begin{equation}
  E\sim A^{3}\,.
\end{equation}
If we compare to our result that the dilation scales as $\Xi\sim A$ (at fixed $H$, since $\Xi\propto\phi$ and $\phi\propto A$), and that the total energy is the energy density $\propto\Xi^2$ times the area, $E\sim A\,\Xi^2$, we find $E\sim A^3$ -- in agreement with the super-extensive scaling.

\subsection{Geometry and invagination energy landscape\label{methods:geometric calculations}}

\subsubsection*{Geometry calculations}

A spherical cap is parameterized by its mean curvature $H = 1/R$ and its closing angle
$\vartheta \in [0,\pi]$.
The surface area of the cap is
\begin{equation}
  A(H,\vartheta) = \frac{2\pi}{H^2}\!\left(1 - \cos\vartheta\right).
  \label{eq:cap_area}
\end{equation}
The \emph{closure} $\phi$ measures the fraction of a full sphere covered by the cap,
\begin{equation}
  \phi \;=\; \frac{AH^2}{4\pi} \;=\; \frac{1-\cos\vartheta}{2} \;\in\; [0,1].
  \label{eq:closure}
\end{equation}
For the surface-tension contribution to the invagination energy
(Sec.~\ref{methods:invagination landscape}), we need the excess area relative to the
flat disc of the same projected radius,
\begin{equation}
  \Delta A \;=\; A - \pi R^2\sin^2\!\vartheta
           \;=\; A\,\frac{1-\cos\vartheta}{2}
           \;=\; \frac{A^2H^2}{4\pi}.
  \label{eq:excess_area}
\end{equation}

\subsubsection*{Invagination energy landscape\label{methods:invagination landscape}}

In the main text we use a reduced two-variable description of the invagination
 process. The state of the pit is described by the membrane area $A
 $ covered by the coat and by the mean curvature $H$ of the corresponding
  spherical cap. This landscape is not meant to be the microscopic simulation Hamiltonian.
   Instead, it is an effective coarse-grained energy landscape in which microscopic lattice degrees of
    freedom, membrane shape modes beyond the spherical-cap approximation, and local relaxation processes have been
     absorbed into a small number of effective parameters. Within this reduced description, 
the invagination is described by the systems state $s=(H, A)$ which changes according to the 
underlying energy landscape. The ground for that is an a-priori \textit{static} energy
landscape $E(H, A, s)$, which depends on the curvature of the system, its
area and changeable system parameters, like the clathrin bending
rigidity or the preferred curvature.

In the spherical cap model, the accessible domain is the region in
the $(H, A)$ plane with closure $\phi\leq 1$, which corresponds to the 
geometric constraint that a spherical cap of
 curvature $H$ cannot have an area larger than
  $A_{\rm max}(H)=4\pi/H^2$.

\begin{equation}
    0\leq \phi=\frac{AH^2}{4\pi}\leq 1,
    \qquad
    A\leq A_{\rm max}(H)=\frac{4\pi}{H^2}.
    \label{eq:landscape_accessible_domain}
\end{equation}
The $\phi=1$ boundary will be essential for the understanding of the
state trajectories in the energy landscape. We will show that the
energy minimum in this landscape, the \textit{attractor}, moves
during maturation of the coat, which is what determines the
shape of the state-trajectory through the $(H,A)$ plane. The state
is "chasing" its attractor. If the attractor is "caught" inside the
accessible state space, the pit becomes aborted. If the attractor moves outside
the accessible region before being caught, the pit closes.

The landscape contains four contributions: membrane bending, 
membrane surface tension, clathrin polymerization, and clathrin bending. 
These terms should be read as an instantaneous effective description of a coat at a given stage of assembly. 
In particular, the clathrin parameters introduced below are the effective parameters 
generated by the lattice state discussed in the previous sections. 
A well established result in the literature is that a spherical closed
membrane has bending energy $4\pi\kappa_{\rm M}$, irrespective of its size (depending
on convention, a factor of $2$ might be different). We will
use this energy scale to normalize all energy contributions to work
in dimensionless units. We write
\begin{equation}
    \mathcal{E}(H,A):=\frac{E(H,A)}{4\pi\kappa_{\rm M}}.
\end{equation}
Here $\kappa_{\rm M}$ is the membrane bending rigidity. With the curvature convention used throughout this supplement, the membrane bending term is
\begin{equation}
    E_{\rm bend}=\kappa_{\rm M}H^2A,
    \qquad
    \mathcal{E}_{\rm bend}=\frac{AH^2}{4\pi}=\phi.
    \label{eq:landscape_membrane_bending}
\end{equation}
Thus membrane bending penalizes curvature and contributes exactly the closure coordinate in the normalized landscape.

The membrane tension term is the work required to create excess membrane area relative to the flat projected disc. Using Eq.~(\ref{eq:excess_area}),
\begin{equation}
    E_{\rm tens}=\sigma\Delta A
    =\sigma\frac{A^2H^2}{4\pi},
    \qquad
    \mathcal{E}_{\rm tens}
    =\frac{\sigma}{\kappa_{\rm M}}\frac{A^2H^2}{(4\pi)^2}.
    \label{eq:landscape_tension}
\end{equation}
Both membrane bending and membrane tension are 
minimized at the origin of the $(H,A)$ plane and 
therefore oppose invagination.

The clathrin polymerization term accounts for the 
free-energy gain of adding triskelia to the coat. 
If $\Delta E_{\rm pol}>0$ denotes the average free-energy 
gain per incorporated triskelion and $A_{\rm cl}$ is 
the average area per triskelion in the assembled 
lattice, then
\begin{equation}
    E_{\rm pol}=-\Delta E_{\rm pol}\frac{A}{A_{\rm cl}},
    \qquad
    \mathcal{E}_{\rm pol}
    =-\frac{\Delta E_{\rm pol}}{4\pi\kappa_{\rm M}}\frac{A}{A_{\rm cl}}.
    \label{eq:landscape_polymerisation}
\end{equation}
This term favors increasing coat area, driving the system
away from the origin. In contrast to the stiffness calculation above, the factor $A_{\rm cl}$ is used here only to convert coat area into an approximate number of incorporated triskelia.

The final term is the clathrin bending energy. Here we explicitly use the effective bending rigidity and effective preferred curvature of the clathrin coat. To keep this distinction visible in the equations, we denote them by $\kappa_{\rm cl}^{\rm eff}$ and $H_{\rm min}^{\rm eff}$,
\begin{equation}
    E_{\rm cl}=\kappa_{\rm cl}^{\rm eff}(H-H_{\rm min}^{\rm eff})^2A,
    \qquad
    \mathcal{E}_{\rm cl}
    =\frac{\kappa_{\rm cl}^{\rm eff}}{4\pi\kappa_{\rm M}}(H-H_{\rm min}^{\rm eff})^2A.
    \label{eq:landscape_clathrin_bending}
\end{equation}
This term favors curvatures close to $H_{\rm min}^{\rm eff}$ and becomes more important as the coat grows. The quantities $\kappa_{\rm cl}^{\rm eff}$ and $H_{\rm min}^{\rm eff}$ are coarse-grained, state-dependent quantities. They correspond to the effective coat rigidity and effective energy minimizing curvature derived above, evaluated for the current lattice state and growth history. In the main text, varying $\kappa_{\rm cl}^{\rm eff}$ is a compact way to show how the stiffening mechanism shifts the energy basin. For analytic clarity we treat $\kappa_{\rm cl}^{\rm eff}$ and $H_{\rm min}^{\rm eff}$ as fixed parameters when minimizing a single instantaneous landscape.

Combining the four terms gives
\begin{equation}
\begin{split}
    \mathcal{E}(H,A)=&\frac{AH^2}{4\pi}
    +\frac{\sigma}{\kappa_{\rm M}}\frac{A^2H^2}{(4\pi)^2}
    +\frac{\kappa_{\rm cl}^{\rm eff}}{4\pi\kappa_{\rm M}}(H-H_{\rm min}^{\rm eff})^2A\\
    &-\frac{\Delta E_{\rm pol}}{4\pi\kappa_{\rm M}}\frac{A}{A_{\rm cl}}.
\end{split}
\label{eq:landscape_total_energy}
\end{equation}
This is the energy landscape shown in Fig.~5 of the main text. When the coat stiffens during assembly, Eq.~(\ref{eq:landscape_total_energy}) should be interpreted quasi-statically: at each stage the current lattice connectivity and growth history determine the effective values of $\kappa_{\rm cl}^{\rm eff}$ and $H_{\rm min}^{\rm eff}$, and the landscape describes the direction in which the coarse variables $H$ and $A$ are biased at that stage.

For a fixed coat area $A$ and fixed instantaneous values of $\kappa_{\rm cl}^{\rm eff}$ and $H_{\rm min}^{\rm eff}$, the curvature of the local energy minimum is obtained from $\partial_H\mathcal{E}=0$. Differentiating Eq.~(\ref{eq:landscape_total_energy}) gives
\begin{equation}
    \partial_H\mathcal{E}
    =\frac{AH}{2\pi}
    +\frac{\sigma}{\kappa_{\rm M}}\frac{A^2H}{8\pi^2}
    +\frac{\kappa_{\rm cl}^{\rm eff}}{2\pi\kappa_{\rm M}}A(H-H_{\rm min}^{\rm eff}).
    \label{eq:landscape_dEdH}
\end{equation}
Solving for $H$ gives the preferred landscape curvature $H^*$:
\begin{equation}
    \boxed{H^*(A)=
    \frac{4\pi\kappa_{\rm cl}^{\rm eff}H_{\rm min}^{\rm eff}}{A\sigma+4\pi(\kappa_{\rm cl}^{\rm eff}+\kappa_{\rm M})}.}
    \label{eq:landscape_Hstar}
\end{equation}
Increasing membrane tension shifts this trajectory to lower curvature, while increasing effective clathrin stiffness shifts it toward $H_{0,\rm cl}^{\rm eff}$.

Growth however is mainly controlled through the coat area $A$. 
Growth is only limited if an energy minimum exists at a finite
area. By plugging in the energy minimising curvature $H^*$ into the
combined energy, we reach an effective one-dimensional energy
\begin{equation}
    \mathcal{E}^*(A):=\mathcal{E}(H^*(A),A)
\end{equation}
and can check if it has a stationary point at positive area. 
The resulting critical area (discarding the negative area solution) is
\begin{equation}
\boxed{
    A_{\rm crit}=\frac{4\pi\kappa_{\rm M}}{\sigma}
    \left[
    \frac{H_{\rm min}^{\rm eff}(\kappa_{\rm cl}^{\rm eff}/\kappa_{\rm M})\sqrt{A_{\rm cl}}\sqrt{\kappa_{\rm M}+\kappa_{\rm cl}^{\rm eff}}}
    {\sqrt{A_{\rm cl}(H_{\rm min}^{\rm eff})^2\kappa_{\rm cl}^{\rm eff}-\Delta E_{\rm pol}}}
    -\left(1+\frac{\kappa_{\rm cl}^{\rm eff}}{\kappa_{\rm M}}\right)
    \right].}
    \label{eq:landscape_Acrit}
\end{equation}
The corresponding curvature at that critical area is
\begin{equation}
    \boxed{H_{\rm crit}=\frac{\sqrt{A_{\rm cl}(H_{\rm min}^{\rm eff})^2\kappa_{\rm cl}^{\rm eff}-\Delta E_{\rm pol}}}
    {\sqrt{A_{\rm cl}}\sqrt{\kappa_{\rm M}+\kappa_{\rm cl}^{\rm eff}}}.}
    \label{eq:landscape_Hcrit}
\end{equation}
Equivalently, with the membrane tension length $\ell_\sigma^2:=\kappa_{\rm M}/\sigma$, Eq.~(\ref{eq:landscape_Acrit}) can be written as
\begin{equation}
    A_{\rm crit}=4\pi\ell_\sigma^2
    \left[
    \frac{\kappa_{\rm cl}^{\rm eff}}{\kappa_{\rm M}}
    \left(\frac{H_{\rm min}^{\rm eff}}{H_{\rm crit}}-1\right)-1
    \right].
    \label{eq:landscape_Acrit_compact}
\end{equation}

Equations~(\ref{eq:landscape_Acrit}) and (\ref{eq:landscape_Hcrit}) 
give a criterion for when the landscape contains an energy basin. 
Such a basin exists only if $H_{\rm crit}$ exists, is real, and $A_{\rm crit}$ is positive.

The critical curvature is real only if
\begin{equation}
    \Delta E_{\rm pol}<\Delta E_{\rm max}:=A_{\rm cl}(H_{\rm min}^{\rm eff})^2\kappa_{\rm cl}^{\rm eff}.
    \label{eq:landscape_DeltaEmax}
\end{equation}
The critical area is positive only if the polymerization gain is large enough,
\begin{equation}
    \Delta E_{\rm pol}>\Delta E_{\rm min}:=
    \frac{A_{\rm cl}(H_{\rm min}^{\rm eff})^2\kappa_{\rm cl}^{\rm eff}}{1+\kappa_{\rm cl}^{\rm eff}/\kappa_{\rm M}}.
    \label{eq:landscape_DeltaEmin}
\end{equation}
Thus a finite basin in the accessible instantaneous landscape requires
\begin{equation}
    \Delta E_{\rm min}<\Delta E_{\rm pol}<\Delta E_{\rm max}.
    \label{eq:landscape_basin_window}
\end{equation}
If polymerization is too weak, coat growth is not favorable. If polymerization is too strong relative to clathrin bending stiffness, the energy decreases by adding more clathrin and no finite-area basin arrests growth. Increasing the effective stiffness $\kappa_{\rm cl}^{\rm eff}$ raises the upper threshold $\Delta E_{\rm max}$ and can therefore create a finite energy basin for a fixed polymerization gain. This is the continuum landscape representation of the stiffening mechanism: as the coat becomes effectively stiffer, and as its effective preferred curvature changes through $H_{\rm min}^{\rm eff}$, the energy minimum moves through the $(H,A)$ plane. If that minimum remains inside the accessible region, the pit can arrest as an aborted pit. If the minimum is displaced beyond the closure boundary $\phi=1$, the accessible landscape drives the coat toward vesicle closure. These conditions should not be read as sharp microscopic fate boundaries, but as a coarse-grained explanation for how changes in membrane tension, polymerization gain, connectivity, and lattice stiffening reorganize the invagination landscape.

\subsection{Coat Cutting}
The same notation gives a compact description of in-silico cutting. A cut
reduces the mechanical contribution of selected rings or bonds. We represent
this by replacing
\begin{equation}
    w_i\rightarrow c_iw_i,
    \qquad 0\le c_i\le 1,
    \label{eq:cut_weight_reduction_current_phi}
\end{equation}
where $c_i=1$ leaves ring $i$ unchanged and $c_i=0$ removes its contribution to
the connected stress-transmitting coat. After cutting,
\begin{equation}
    \mathcal{W}_{\rm after}=\sum_i c_iw_i,
    \qquad
    \mathcal{H}_{\rm after}
    =\frac{\sum_i c_iw_iH_i}{\sum_i c_iw_i}.
    \label{eq:cut_WH_after_current_phi}
\end{equation}
The post-cut effective parameters are therefore
\begin{equation}
    \kappa_{{\rm cl},\,{\rm after}}^{\rm eff}
    =\kappa_{\rm cl}(1+\mathcal{W}_{\rm after})
    \label{eq:cut_kappa_after_current_phi}
\end{equation}
and
\begin{equation}
    H_{{\rm min},\,{\rm after}}^{\rm eff}
    =\frac{H_{\rm min}^{\rm micro}+\mathcal{W}_{\rm after}\mathcal{H}_{\rm after}}
    {1+\mathcal{W}_{\rm after}}.
    \label{eq:cut_Hpref_after_current_phi}
\end{equation}

We define the positive removed weight as
\begin{equation}
    \delta\mathcal{W}
    :=\mathcal{W}-\mathcal{W}_{\rm after}
    =\sum_i(1-c_i)w_i>0.
    \label{eq:cut_removed_WH_current_phi}
\end{equation}
We will see that the material removed by the cut carries the weighted memorised curvature
\begin{equation}
    H_{\rm cut}
    :=
    \frac{\sum_i(1-c_i)w_iH_i}
    {\sum_i(1-c_i)w_i}
    \label{eq:Hcut_current_phi}
\end{equation}which we therefore define already here.
The post-cut quantities can therefore be written
as
\begin{equation}
    \mathcal{W}_{\rm after}
    =\mathcal{W}-\delta\mathcal{W},
    \qquad
    \mathcal{W}_{\rm after}\mathcal{H}_{\rm after}
    =\mathcal{W}\mathcal{H}-\delta\mathcal{W}H_{\rm cut}.
    \label{eq:cut_after_memory_balance_current_phi}
\end{equation}
Within the linearized stiffness expression, cutting lowers the effective
bending rigidity by
\begin{equation}\boxed{
    \Delta\kappa_{\rm cl}^{\rm eff}
    =
    \kappa_{{\rm cl},\,{\rm after}}^{\rm eff}
    -\kappa_{\rm cl}^{\rm eff}
    =
    -\kappa_{\rm cl}\,\delta\mathcal{W}<0.
    \label{eq:cut_delta_kappa_current_phi}}
\end{equation}
For the preferred curvature, inserting
Eq.~(\ref{eq:cut_after_memory_balance_current_phi}) into
Eq.~(\ref{eq:cut_Hpref_after_current_phi}) gives
\begin{equation}
    H_{{\rm min},\,{\rm after}}^{\rm eff}
    =
    \frac{H_{\rm min}^{\rm micro}+\mathcal{W}\mathcal{H}
    -\delta\mathcal{W}H_{\rm cut}}
    {1+\mathcal{W}-\delta\mathcal{W}}.
    \label{eq:cut_Hpref_after_balance_current_phi}
\end{equation}
Subtracting the intact value
$H_{\rm min}^{\rm eff}=(H_{\rm min}^{\rm micro}+\mathcal{W}\mathcal{H})/(1+\mathcal{W})$ gives
the exact finite-cut expression within the same linearized model,
\begin{equation}
\begin{split}
    \Delta H_{\rm min}^{\rm eff}
    &=
    H_{{\rm min},\,{\rm after}}^{\rm eff}
    -H_{\rm min}^{\rm eff}\\
    &=
    \frac{\delta\mathcal{W}}
    {1+\mathcal{W}-\delta\mathcal{W}}
    \left(H_{\rm min}^{\rm eff}-H_{\rm cut}\right).
\end{split}
    \label{eq:cut_delta_Hpref_intermediate_current_phi}
\end{equation}
For a weak cut, $\delta\mathcal{W}\ll1+\mathcal{W}$, this reduces to
\begin{equation}
    \boxed{
    \Delta H_{\rm min}^{\rm eff}
    \simeq
    \frac{\delta\mathcal{W}}{1+\mathcal{W}}
    \left(H_{\rm min}^{\rm eff}-H_{\rm cut}\right).}
    \label{eq:cut_delta_Hpref_current_phi}
\end{equation}
Cutting increases the effective preferred curvature if the removed material
carries a lower memorized curvature than the intact coat,
$H_{\rm cut}<H_{\rm min}^{\rm eff}$. This is the typical situation for cuts
that remove early material incorporated at low curvature. In that case, the cut
both lowers the effective rigidity and erases part of the coat's low-curvature
memory, allowing the relaxed coat to shift toward higher preferred curvature.

\section{Simulation code and reproducibility workflow\label{sec:Simulations}}

The numerical model used in this work is implemented as a Python/JAX codebase
that accompanies the manuscript. The repository is organised as a script-based
research workflow rather than as an installable Python package. Its purpose is
to make the simulations, post-processing steps, and figure-generation scripts
transparent and reproducible. The public release contains the active simulation
and analysis pipeline, a minimal runnable example, and the scripts used to
assemble the paper figures. Local development archives, exploratory notebooks,
large simulation outputs, and generated figures are intentionally excluded from
the tracked release repository.

\subsection{Repository organisation}

The core simulation code is located in \texttt{simulation\_code/}. The active
entry point for running simulations is \texttt{run\_simulation\_batch.py},
which reads a parameter table and dispatches individual simulations to
\texttt{simulation\_worker.py}. The worker constructs a \texttt{Grid} object,
runs the kinetic Monte Carlo evolution, and writes the resulting trajectory
files. The main model classes and geometry helpers are in
\texttt{simulation\_code/classes/}. The most important files are:

\begin{itemize}
  \item \texttt{classes/grid.py}: the \texttt{Grid} class, which owns the full
  lattice state, membrane geometry, random number generator, kinetic Monte
  Carlo event logic, energy calculations, local relaxation, and plotting
  helpers.

  \item \texttt{classes/node.py}: the \texttt{Node} class, representing one
  clathrin triskelion hub with a position, activity state, up to three direct
  binding partners, and next-nearest-neighborlinks.

  \item \texttt{classes/projection.py} and
  \texttt{classes/helper\_methods.py}: geometric and numerical helper routines,
  including stereographic projection between the sphere and local coordinate
  charts.

  \item \texttt{classes/face\_counting\_method.py}: graph-based detection of
  polygonal motifs such as pentagons, hexagons, and heptagons in the clathrin
  lattice.

  \item \texttt{calculate\_bending\_rigidity.py}: post-processing script for
  extracting an effective coat rigidity and preferred curvature from saved
  configurations by curvature sweeps and energy fitting.

  \item \texttt{cutting\_experiment.py} and
  \texttt{cutting\_code\_polished.py}: scripts and helper functions for
  geodesic cutting experiments, topology checks, post-cut relaxation, and
  before/after rigidity measurements.

  \item \path{consolidate_sim_results.py} and
  \path{consolidate_cutexp_results.py}: utilities that combine many
  individual simulation or cutting-experiment folders into compact dictionaries
  used by the figure-generation scripts.
\end{itemize}

The repository also contains an \texttt{examples/} folder. These examples are
not intended to reproduce the full production data set. Instead, they are small
end-to-end tests that demonstrate the file layout and show how a new user can
run a simulation, read the output, calculate rigidity values, perform a small
cutting experiment, consolidate results, and export frames for Blender-based
visualization. The paper figure scripts are stored in \texttt{figures/}. They
operate on consolidated output files generated from the full simulation data
sets, which are too large to be stored in the public code repository.

\subsection{Representation of the clathrin coat}

The coat is represented as a growing graph embedded on a spherical membrane
patch. Each active node corresponds to one clathrin hub. Its projected
coordinates are stored in a local stereographic chart, while helper functions
convert between projected coordinates and three-dimensional points on the
sphere whenever geometric distances, tangent vectors, or visualisations are
needed. The node state also stores direct nearest-neighborpartners and
next-nearest-neighborrelationships. Unoccupied partner slots are represented
by sentinel values, so that the node arrays have fixed shapes and can be handled
inside JAX-compiled functions.

The \texttt{Grid} object stores all global parameters and dynamic state:
membrane curvature, preferred curvature, membrane bending rigidity, surface
tension, microscopic spring constants, polymerization energies, the random
number generator key, the batched node object, and auxiliary matrices such as
active-node distance information. Both \texttt{Grid} and \texttt{Node} are
registered as JAX pytrees. Their pytree definitions separate dynamic JAX arrays,
which can be traced by JIT-compiled functions, from static Python-side
configuration such as node counts, spring constants, and algorithmic settings.
This organization is essential for compiling the kinetic Monte Carlo loop while
keeping the code close to the physical object model.

\subsection{Kinetic Monte Carlo evolution}

The central evolution routine is \texttt{Grid.evolve\_grid\_til}. It is
implemented with \texttt{jax.lax.scan}, so that a complete trajectory of a fixed
number of kinetic Monte Carlo steps can be compiled and executed efficiently.
At each step, the code constructs the set of currently possible microscopic
events, samples one event from the event catalogue, evaluates its energy change,
and accepts or rejects the event using a rate- and energy-dependent Metropolis
criterion.

The current event catalogue contains five classes of moves: addition of a new
node bonded to an available leg of an existing node, formation of a direct
nearest-neighborbond, formation of a next-nearest-neighborbond, removal of a
direct bond, and removal of a next-nearest-neighborbond. After an accepted or
rejected lattice event, the kinetic Monte Carlo time is advanced according to
the total event rate. The grid then performs local relaxation moves. In the
variable-curvature simulations, the sphere radius can change adiabatically,
allowing the membrane patch to invaginate as the lattice grows.

During a batch run, the command-line runner can print controlled progress
messages. For example, \texttt{--verbose} reports every evolution step in the
small example, while \texttt{--progress-interval 100} reports only every 100
steps for longer production runs. The first run for a new parameter shape can be
slow because JAX compiles the traced computation before executing it.

\subsection{Energy terms and recorded observables}

The microscopic energy combines membrane and lattice contributions. The
membrane part contains a Helfrich bending term and a surface-tension term. The
lattice part contains harmonic contributions for bond stretching, in-plane leg
angles, out-of-plane leg angles, and next-nearest-neighborgeometry, together
with polymerization terms for leg and hub binding. Steric exclusion terms avoid
unphysical overlap of nodes. The parameters controlling these contributions are
read from a spreadsheet or CSV file by the batch runner.

For each simulation, the code stores a time series of observables in
\texttt{data\_dict.pkl}. The most important entries are the kinetic Monte Carlo
time, the sampled process class, the acceptance flag, total energy, active node
number, direct and next-nearest-neighborbond numbers, polygon counts,
individual energy contributions, acceptance probability, sphere radius,
curvature, invagination angle, and membrane area. The final \texttt{Grid} object
is saved in \texttt{grid.pkl}, and the full time-stacked node trajectory is saved
in \texttt{time\_stacked\_batched\_nodes.pkl}. A human-readable
\texttt{parameters.txt} file records the parameter row and random seed for each
replicate.

The standard output folder for one simulation therefore contains
\begin{verbatim}
parameters.txt
grid.pkl
data_dict.pkl
time_stacked_batched_nodes.pkl
\end{verbatim}
with an additional \texttt{calculated\_data\_dict.pkl} file if post-processing
quantities such as bending rigidity have been calculated.

\subsection{Bending-rigidity and cutting post-processing}

The effective coat rigidity is computed from saved coat configurations. For a
chosen timestep, \texttt{calculate\_bending\_rigidity.py} loads the grid and the
stored node configuration, imposes a set of nearby curvatures, relaxes internal
node coordinates at each curvature, and fits the resulting energy--curvature
response. The default output is stored under the key
\texttt{bending\_rigidity} in \texttt{calculated\_data\_dict.pkl}. The same
calculation is used inside the cutting workflow to compare the mechanical
response before and after a geodesic cut.

The cutting experiment loads a saved simulation folder and defines a geodesic
arc on the spherical membrane. Nodes within a specified arc-distance of this
geodesic are removed, small disconnected fragments can be discarded, and the
remaining coat is relaxed. The workflow records the cut geometry, removed node
indices, component sizes, topology-consistency checks, curvature before and
after relaxation, and rigidity estimates before and after the cut. This provides
the computational basis for perturbing assembled coats and measuring how local
connectivity affects the global curvature response.

\subsection{Consolidation and figure generation}

Large parameter sweeps produce one folder per replicate simulation. To make
figure generation independent of the raw folder hierarchy, the helper script
\texttt{consolidate\_sim\_results.py} scans a results directory and writes a
single \texttt{consolidated\_sim\_results.pkl} file. This file contains three
top-level entries: \texttt{summary}, a compact list of final observables and
parameters for each run; \texttt{results}, a dictionary containing the full
\texttt{data\_dict}, optional \texttt{calculated\_data\_dict}, parameters, and
source path for each run; and \texttt{metadata}, which records the source
directory and creation information. The paper figure scripts read these
consolidated files rather than scanning raw simulation folders directly.

The public repository includes the figure-generation scripts and small
figure-local assets. The full simulation output archive is not tracked because
it is large and machine-specific. Consequently, several figure scripts contain
local data-path variables that point to the production data archive used for the
manuscript. To rerun the figures on another machine, these path variables must
be replaced by the corresponding local paths to the consolidated simulation or
cutting-experiment results. This separation keeps the code release lightweight
while preserving the exact analysis scripts used for the paper figures.

\subsection{Reproducibility checks and example workflows}

The release repository includes a minimal example that can be run after creating
the conda environment specified by \texttt{environment.yaml}. The minimal
workflow executes a short simulation from a CSV parameter table, writes the
standard output files, and plots basic time series from \texttt{data\_dict.pkl}.
Additional examples show how to run a small rigidity calculation, perform a
small cutting experiment, consolidate simulation outputs, and export trajectory
frames to Blender-friendly \texttt{.npz} files. These examples are intentionally
small. Their purpose is to verify that the code, file formats, and analysis
interfaces work in a fresh checkout; the production simulations used in the
paper use larger parameter tables, longer trajectories, and more expensive
post-processing settings.

For publication, the repository was tested with a fresh local clone: the conda
environment was created from \texttt{environment.yaml}, the minimal simulation
was run from \path{examples/01_minimal_simulation/}, the output was analyzed,
and the resulting folder was consolidated successfully. Generated output files,
JAX compilation caches, local archives, and large data products are ignored by
Git. The code is released under the MIT License to permit reuse, modification,
and redistribution with attribution.

\clearpage

\paragraph{Caption for Movie S1.}
{Kinetic Monte Carlo simulation of clathrin assembly at fixed membrane curvature of $H=$~0.35~$L_0^{-1}$. Nodes are colored by their energy from blue (low) to red (high), while bonds are color coded by the number of leg segments coinciding on that bond, from 2 (blue) over 3 (yellow, one next nearest neighbor bond) to 4 (red, two next nearest neighbor bonds).}

\paragraph{Caption for Movie S2.}
{Kinetic Monte Carlo simulation of clathrin assembly at fixed membrane curvature of $H=$~0.2~$L_0^{-1}$. Nodes are colored by their energy from blue (low) to red (high), while bonds are color coded by the number of leg segments coinciding on that bond, from 2 (blue) over 3 (yellow, one next nearest neighbor bond) to 4 (red, two next nearest neighbor bonds).}

\paragraph{Caption for Movie S3.}
{Kinetic Monte Carlo simulation of clathrin assembly at variable membrane curvature with surface tension ratio $\sigma/\kappa_M=$~0.4, assembling into a flat plaque. Nodes are colored by their energy from blue (low) to red (high), while bonds are color coded by the number of leg segments coinciding on that bond, from 2 (blue) over 3 (yellow, one next nearest neighbor bond) to 4 (red, two next nearest neighbor bonds).}

\paragraph{Caption for Movie S4.}
{Kinetic Monte Carlo simulation of clathrin assembly at variable membrane curvature with surface tension of ratio $\sigma/\kappa_M=$~0.2, assembling into an aborted pit. Nodes are colored by their energy from blue (low) to red (high), while bonds are color coded by the number of leg segments coinciding on that bond, from 2 (blue) over 3 (yellow, one next nearest neighbor bond) to 4 (red, two next nearest neighbor bonds).}

\paragraph{Caption for Movie S5.}
{Kinetic Monte Carlo simulation of clathrin assembly at variable membrane curvature with surface tension ratio $\sigma/\kappa_M=$~0.025, assembling into a closed cage. Nodes are colored by their energy from blue (low) to red (high), while bonds are color coded by the number of leg segments coinciding on that bond, from 2 (blue) over 3 (yellow, one next nearest neighbor bond) to 4 (red, two next nearest neighbor bonds).}

\paragraph{Caption for Movie S6.}
{Animation of in silico cutting experiment. A previously simulated cage at surface tension ratio $\sigma/\kappa_M=$~0.05, is cut along a geodesic. Afterwards we perform 120,000 Monte Carlo relaxation steps, with the cage increasing its curvature. Bond and node colors encode the
  local energetic load on a logarithmic scale (blue is low, red is high).
}

\bibliography{literature}
\bibliographystyle{abbrv}